\documentclass[twocolumn]{IEEEtran}
\usepackage{graphicx}
\usepackage{psfrag}
\usepackage[tight,footnotesize]{subfigure}
\usepackage{amsmath}
\usepackage{amssymb}
\usepackage{mathrsfs}
\usepackage{stfloats}
\usepackage{dsfont}
\usepackage{setspace}
\usepackage{array}
\usepackage{bbm}
\usepackage{mathabx}
\usepackage{algorithmic}
\usepackage{algorithm}
\usepackage{glossaries}
\usepackage{color}

\newcommand{\eeq}{\end{equation}}

\newcommand{\bq}{\mbox{\boldmath $q$}}

\newcommand{\bA}{\mbox{\boldmath $A$}}

\newcommand{\bh}{\mbox{\boldmath $h$}}
\newcommand{\bH}{\mbox{\boldmath $H$}}

\newcommand{\bG}{\mbox{\boldmath $G$}}
\newcommand{\bPhi}{\mbox{\boldmath $\Phi$}}

\newcommand{\bu}{\mbox{\boldmath $u$}}

\newcommand{\bg}{\mbox{\boldmath $g$}}

\newcommand{\bv}{\mbox{\boldmath $v$}}

\newcommand{\bI}{\mbox{\boldmath $I$}}

\newcommand{\ds}{\displaystyle}
\newcommand{\bw}{\mbox{\boldmath $w$}}

\newcommand{\beq}{\begin{equation}}

\newtheorem{lemma}{Lemma}
\newtheorem{proposition}{Proposition}

\makeglossaries

\newacronym{mac}{MAC}{multiple-access channel}
\newacronym{bc}{BC}{broadcast channel}
\newacronym{mimo}{MIMO}{multiple-input multiple-output}
\newacronym{siso}{SISO}{single-input single-output}
\newacronym{sc}{SC}{single-carrier}
\newacronym{mc}{MC}{multi-carrier}
\newacronym{ofdma}{OFDMA}{orthogonal frequency division multiple access}
\newacronym{af}{AF}{amplify-and-forward}
\newacronym{df}{DF}{decode-and-forward}
\newacronym{cf}{CF}{compress-and-forward}
\newacronym{mwrc}{MWRC}{multi-way relay channel}
\newacronym{pde}{PDE}{partial data exchange}
\newacronym{fde}{FDE}{full data exchange}
\newacronym{iid}{i.i.d.\@}{independent and identically distributed}
\newacronym{awgn}{AWGN}{additive white Gaussian noise}
\newacronym{awg}{AWG}{additive white Gaussian}
\newacronym{sic}{SIC}{successive interference cancellation}
\newacronym{snr}{SNR}{signal-to-noise ratio}
\newacronym{sinr}{SINR}{signal-to-interference-plus-noise ratio}
\newacronym{ber}{BER}{bit error rate}
\newacronym{zf}{ZF}{zero-forcing}
\newacronym{mmse}{MMSE}{minimum mean square error}
\newacronym{sud}{SUD}{single user decoding}
\newacronym{dof}{DoF}{degrees of freedom}
\newacronym{gdof}{GDoF}{generalized degrees of freedom}
\newacronym{nnc}{NNC}{noisy network coding}
\newacronym{dmn}{DMN}{discrete memoryless network}
\newacronym{csi}{CSI}{channel state information}
\newacronym{ee}{EE}{energy efficiency}
\newacronym{ian}{IAN}{treating interference as noise}
\newacronym{snd}{SND}{simultaneous non-unique decoding}
\newacronym{brd}{BRD}{best response dynamics}
\newacronym{br}{BR}{best response}
\newacronym{ne}{NE}{Nash equilibrium}
\newacronym{lhs}{LHS}{left-hand side}
\newacronym{rhs}{RHS}{right-hand side}
\newacronym{gee}{GEE}{global energy efficiency}
\newacronym{wsee}{WSEE}{weighted sum energy efficiency}
\newacronym{wpee}{WPEE}{weighted product energy efficiency}
\newacronym{wmee}{WMEE}{weighted minimum energy efficiency}
\newacronym{kkt}{KKT}{Karush-Kuhn-Tucker}
\newacronym{pc}{PC}{pseudo-concave}
\newacronym{qc}{QC}{quasi-concave}
\newacronym{ql}{QL}{quasi-linear}
\newacronym{pl}{PL}{pseudo-linear}
\newacronym{spc}{SPC}{strictly pseudo-concave}
\newacronym{sqc}{SQC}{strictly quasi-concave}
\newacronym{lfp}{LFP}{linear fractional problem}
\newacronym{clfp}{CLFP}{concave-linear fractional problem}
\newacronym{ccfp}{CCFP}{concave-convex fractional problem}
\newacronym{mmfp}{MMFP}{max-min fractional problem}
\newacronym{sorp}{SoRP}{sum-of-ratios problem}
\newacronym{porp}{PoRP}{product-of-ratios problem}
\newacronym{srp}{SRP}{single-ratio problem}
\newacronym{brb}{BRB}{branch-reduce-and-bound}
\newacronym{qos}{QoS}{quality-of-service}
\newacronym{comp}{CoMP}{cooperative multi-point}
\newacronym{ue}{UE}{user equipment}
\newacronym{bs}{BS}{base station}
\newacronym{mrc}{MRC}{maximum ratio combining}
\newacronym{d2d}{D2D}{device-to-device}
\newacronym{lmmse}{LMMSE}{linear minimum mean square error}
\newacronym{ris}{RIS}{reconfigurable intelligent surface}
\newacronym{svd}{SVD}{singular values decomposition}

\usepackage{accents}

\usepackage[centerlast,small]{caption}

\begin{document}
\title{\huge Overhead-Aware Design of Reconfigurable Intelligent Surfaces in Smart Radio Environments}

\author{Alessio Zappone,~\IEEEmembership{Senior Member~IEEE}, Marco Di Renzo,~\IEEEmembership{Fellow~IEEE}, Farshad Shams,~\IEEEmembership{Member,~IEEE}, Xuewen Qian, Merouane Debbah,~\IEEEmembership{Fellow~IEEE}
\thanks{A. Zappone is with the University of Cassino and Southern Lazio, Cassino, Italy (alessio.zappone@unicas.it). M. Di Renzo, F. Shams, and X. Qian are with Universit\'e Paris-Saclay, CNRS, CentraleSup\'elec, Laboratoire des Signaux et Syst\`emes, 3 Rue Joliot-Curie, 91192 Gif-sur-Yvette, France. (marco.direnzo@centralesupelec.fr). M. Debbah is with Huawei France R\&D, Boulogne-Billancourt, France. This work was supported in part by the European Commission through the H2020 ARIADNE project under grant number 675806 and the H2020 REDESIGN project under grant number 789260.}}

\maketitle

\begin{abstract}
Reconfigurable intelligent surfaces have emerged as a promising technology for future wireless networks. Given that a large number of reflecting elements is typically used and that the surface has no signal processing capabilities, a major challenge is to cope with the overhead that is required to estimate the channel state information and to report the optimized phase shifts to the surface. This issue has not been addressed by previous works, which do not explicitly consider the overhead during the resource allocation phase. This work aims at filling this gap, by developing an overhead-aware resource allocation framework for wireless networks where reconfigurable intelligent surfaces are used to improve the communication performance. An overhead model is proposed and incorporated in the expressions of the system rate and energy efficiency, which are then optimized with respect to the phase shifts of the reconfigurable intelligent surface, the transmit and receive filters, the power and bandwidth used for the communication and feedback phases. The bi-objective maximization of the rate and energy efficiency is investigated, too. The proposed framework characterizes the trade-off between optimized radio resource allocation policies and the related overhead in networks with reconfigurable intelligent surfaces. 
\end{abstract}
\vspace{-0.5cm}
\section{Introduction} \label{Introduction}
Future wireless networks will be a pervasive platform, which will not only connect us but will embrace us through a plethora of services. The ubiquity, speed, and low latency of such networks will allow currently disparate devices and systems to become a distributed intelligent communications, sensing, and computing platform \cite{6G}. Small-cell networks \cite{SmallCells}, massive multiple-input-multiple-output systems \cite{mMIMO}, and millimeter-wave communications \cite{mmWave} are three fundamental technologies that will spearhead the emergence of future wireless networks \cite{5G}. The question is, however, whether these technologies will be sufficient to meet the requirements of future networks that integrate communications, sensing, and computing in a single platform. Wireless networks, in addition, are evolving towards a software-defined paradigm, where every part of the network can be configured and controlled via software \cite{5GPPP_SoftNet}, \cite{NetSlicing}. However, the wireless environment, i.e., the channel, is generally uncontrollable, and often an impediment to be reckoned with, e.g. signal attenuation limits network connectivity, multi-path propagation results in fading, reflections from objects produce uncontrollable interference. 

Motivated by these considerations, the concept of ``smart radio environment'' has recently emerged \cite{TCOM_AI,Pechac_1,SmartWireless,RIS_GE_JSAC}, wherein the environmental objects are envisioned to be coated with man-made intelligent surfaces of configurable electromagnetic materials that are referred to as \glspl{ris} \cite{GeneralizedSnellLaw}, \cite{Metasurfaces_SOTA}. These materials are expected to contain integrated electronic circuits and software that will enable them to control the wireless medium \cite{SmartWireless}, \cite{HyperSurface}. Conceptually, an RIS can be viewed as a reconfigurable mirror or lens, depending on its configuration \cite{DiRenzo2020}, that is made of a number of elementary elements, often referred to as meta-atoms or passive scatterers, that are configurable and programmable in software. The input-output response of each passive scatterer can be appropriately customized, so that the signals impinging upon the \gls{ris} can be predominantly reflected or transmitted in specified directions or focused towards specified locations \cite{PhaseGradientOpt}, \cite{PhaseGradientOptNew}. RISs have the potential to enable the control of the propagation environment, thus potentially changing the design of wireless networks. 

Due to the potential opportunities offered by \gls{ris}-empowered wireless networks, a large body of research contributions have recently appeared in the literature. The interested readers are referred to the survey papers in \cite{SmartWireless,Liaskos_COMMAG,RuiZhang_COMMAG,BasarMag2019,HuangMag2020}, where a comprehensive description of the state-of-the-art, the scientific challenges, the distinctive differences with other technologies, and the open research issues are comprehensively discussed. In \cite{HuLIS2018}, systems made of large active surfaces are put forth as the natural evolution of massive MIMO systems. A similar idea is embraced in \cite{Shlezinger2019}, where it is elaborated on  how \glspl{ris} can be used to implement massive MIMO systems, replacing each conventional antenna with an active reconfigurable surface. The fundamental performance of the system is analyzed, showing that it grants satisfactory performance, while at the same time reducing costs, power consumption, and physical size. In \cite{DiRenzo2019} it is shown how \glspl{ris} can yield better performance compared to the use of relays. Moreover, in \cite{Lu2019} it is shown that RISs can improve the secrecy of communication by focusing the transmit signal only towards the direction of the intended receivers. Recently, in addition, a few experimental testbeds have been built to substantiate the feasibility of \glspl{ris}, e.g., \cite{Fink}, \cite{Caloz_TAP2018}, \cite{Southest_1,Southest_2,Dai2019}. In the following two sub-sections, we describe the contributions that are most related to the present paper, and outline novelty and contributions of our work.

\subsection{Related Works}
We focus our attention on the issue of resource allocation in RIS-empowered wireless networks. In this context, several research papers have appeared recently, mostly considering application scenarios where the line-of-sight link is either too weak or is not available, and, therefore, an RIS is employed to enable the communication through the optimization of the phase shifts of its individual passive elements and of the precoding and decoding vectors of the transmitter and receiver, respectively. 
In \cite{EE_RISs}, the rate and energy efficiency are optimized in \gls{ris}-based multiple input single output (MISO) downlink systems. Alternating optimization of the base station beamformer and of the \gls{ris} phase shifts is performed by means of fractional programming methods for power optimization, and sequential optimization methods for phase optimization. A similar setup is considered in \cite{Wu2018}, with the difference that the problem of power minimization subject to minimum rate constraints is considered. A suboptimal numerical method is proposed based on alternating optimization. In \cite{Yang2019}, a MISO downlink system is analyzed, with the addition that the orthogonal frequency division multiplexing (OFDM) transmission scheme is considered, and the problem of sum-rate maximization is addressed. Sum-rate maximization is also investigated in \cite{Yu2019}, where computationally-efficient, but sub-optimal, algorithms are devised for an \gls{ris}-based MISO system, to optimize  the transmit beamformer and the \gls{ris} phase shifts, still based on the use of alternating optimization. Similarly, alternating optimization methods are used in \cite{Guo2019} to tackle the problem of sum-rate maximization in a MISO downlink system. The base station beamformer and the \gls{ris} phase shifts are optimized, with the additional difficulty that discrete phase-shifts at the \gls{ris} are assumed. In \cite{Jiang2019}, an \gls{ris} is used to boost the performance of over-the-air computations in a multi-user MISO channel. A method based on alternating optimization and difference convex programming is developed, which outperforms semi-definite relaxation alternatives. In \cite{Chen2019}, an \gls{ris} is used to enhance the secrecy rate of a MISO downlink channel with multiple eavesdroppers. Alternating maximization is  used to devise a practical, yet suboptimal, method to optimize the transmit beamformer and the \gls{ris} phase shifts. In \cite{Cui2019}, the use of \glspl{ris} for physical layer security is envisioned, thanks to the possibility of \glspl{ris} to reflect incoming signals towards specified directions. In \cite{Shen2019}, the maximization of the secrecy rate in an \gls{ris}-based multiple-antenna system is investigated, and alternating optimization is used to optimize the transmit beamformer and the \gls{ris} phase shifts. In \cite{Li2019b}, a massive MIMO system is considered, in which multiple RISs equipped with a large number of reflecting elements are deployed and the problem of maximizing the minimum signal-to-interference-plus-noise-ratio at the users is tackled by jointly optimizing the transmit precoding vector and the \glspl{ris} phase shifts. In \cite{Ma2019}, it is shown that the use of \glspl{ris} enhances the performance of systems based on unmanned aerial vehicles (UAVs) upon optimizing the UAV height and various \gls{ris} parameters such as the size, altitude, and distance from the base station. In \cite{Liu2019}, the problem of precoding design in an  \gls{ris}-based multi-user MISO wireless system is addressed, assuming that only discrete phase shifts at the \gls{ris} are possible. The maximization of the rate in an \gls{ris}-assisted MIMO link is tackled in \cite{Ning2019b}, by considering that the \gls{ris} is deployed to assist the communication between the transmitter and the receiver. In \cite{Han2019b}, the problem of power control for physical-layer broadcasting under quality of service constraints for the mobile users is addressed in \gls{ris}-empowered networks. The downlink of a MIMO multi-cell system is considered in \cite{Pan2019b}, where an \gls{ris} is deployed at the boundary between multiple cells. Therein, the  problem of weighted sum-rate maximization is tackled by alternating optimization of the base station beamformer and of the \gls{ris} phase shifts. \gls{ris}-based millimiter wave systems are considered in \cite{Wang2019b}, with reference to a single-user MISO channel. The transmit beamforming and the \gls{ris} phase shifts are optimized considering both the single-\gls{ris} and multi-\gls{ris} cases. In \cite{You2019}, joint channel estimation and sum-rate maximization is tackled in the uplink of a single-user \gls{ris}-based system, where the phase shifts of the \gls{ris} have a discrete resolution. In \cite{PanJSAC2020}, the sum-rate of a MIMO RIS-based system is optimized with respect to the transmitter beamforming and the RIS phase shifts, in the case in which simultaneous information and power transfer is employed.

\subsection{Novelty and Contribution}
The common denominator of all the above works dealing with radio resource allocation is that the optimization is focused only on the data communication phase, whereas the overhead required to estimate the channel state information and to report the optimized phase shifts configuration to the \gls{ris} is not taken into account. As recently highlighted in \cite{SmartWireless}, the overhead for resource allocation in \gls{ris}-empowered wireless networks may be more critical than in conventional wireless networks. This is due to the possibly large number of passive elements in each \gls{ris} that may be spatially distributed throughout the network. Moreover, the above mentioned works optimize the phase shifts of the RISs based on numerical methods, which makes it difficult to assess the ultimate performance of \gls{ris}-empowered wireless networks. 

In contrast, this work develops a resource allocation framework that explicitly accounts for the overhead associated with channel estimation and with the configuration of the optimal \gls{ris} phase shifts. A point-to-point \gls{ris}-based system with multiple antennas at the transmitter and receiver is considered. More precisely, the following specific contributions are made:
\begin{itemize}
\item We propose a model to account for channel estimation and the overhead required for the configuration of the \gls{ris} phase shifts. Based on the overhead-aware expressions of the system rate and energy efficiency, we develop efficient radio resource allocation algorithms. This is a different approach compared to robust resource allocation methods which assume imperfect channel state information \cite{Zhou2020,Hong2020,Zhou2020b}. Indeed, we propose a framework that accounts for the feedback that is necessary for realiable channel estimation and RIS phase shifts deployment, and optimize the system resources based on this new model.
\item We derive two methods for the joint optimization of the \gls{ris} phase shifts, and of the precoding and decoding filters. Both methods are expressed in closed-form, thus requiring a negligible computational complexity compared to state-of-the-art methods based on alternating optimization, as well as enabling  analytical performance evaluation of \gls{ris}-empowered wireless networks. Both approaches are  provably optimal in the case of rank-one channels, which includes the notable special case of single-antenna transmitters and receivers.  
\item We introduce globally optimal algorithms for computing the power and bandwidth that maximize the rate, the energy efficiency, and their trade-off, based on convex/pseudo-convex problems with limited complexity. 
\item Finally, we provide extensive numerical results to show the performance of the proposed approaches. We find that our proposed closed-form phase optimization solution perform similar to more complex, state-of-the-art numerical methods, e.g. alternating optimization. 
\end{itemize}
The rest of the paper is organized as follows. Section \ref{Sec:SysModel} introduces the system model and the problem statement. Section \ref{Sec:PhaseOpt} develops the optimization methods for the allocation of the RIS phase shifts, the beamforming vector, and the receive filter. Section \ref{Sec:ContinuousRate} optimizes the powers and bandwidths for the maximization of the system rate, energy efficiency, and the derivation of the optimal rate-energy trade-off. Section \ref{Sec:Numerics} numerically analyzes the proposed optimization methods. Finally, concluding remarks are given in Section \ref{Sec:Conclusions}. 

\section{System Model and Problem Statement}\label{Sec:SysModel}
\begin{figure}[!h]
\centering
\includegraphics[width=0.3\textwidth]{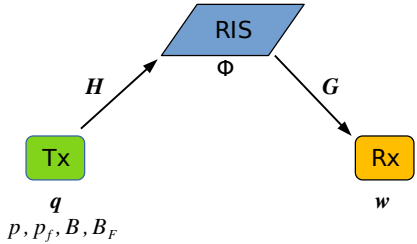}
\caption{System model}
\label{fig:SystemModel}
\end{figure}

The considered system model is depicted in Fig. \ref{fig:SystemModel}.
A transmitter equipped with $N_{T}$ antennas and a receiver equipped with $N_{R}$ antennas communicate through an \gls{ris}. A single-stream transmission is adopted, in order to exploit the diversity gain ensured by the presence of the \gls{ris} and of the multiple transmit and receive antennas\footnote{A more general scenario is represented by a multi-stream transmission, which trades-off reliability with throughput. However, this scenario would lead to more cumbersome expressions of the rate and energy efficiency functions and is left as future work.}. The case under analysis models point-to-point links, but also downlink or uplink communications in cellular networks where multi-user interference is suppressed (for example by means of any orthogonal signaling protocols such as frequency or time division multiple access, or by orthogonal frequency division multiple access).

We assume that no direct link between the transmitter and receiver exists, and we denote by $\bH$ and $\bG$ the channels from the transmitter to the \gls{ris} and from the \gls{ris} to the receiver, respectively, by $\bq$ the unit-norm transmit beamformer, and by $\bw$ the unit-norm receive combiner. Among the different implementations of \glspl{ris} \cite{RIS_GE_JSAC,Qian2020}, we consider surfaces that are made of large arrays of inexpensive
antennas that are spaced half of the wavelength apart and that are individually controlled and tuned. More specifically, we assume that the \gls{ris} is made of $N$ elementary individually and locally optimized passive scatterers, which are capable of independently reflecting the radio wave impinging upon them, by applying a phase shift denoted by $\phi_{n}$, with $n=1,\ldots,N$, which we collect in the diagonal matrix $\bPhi=\text{diag}(e^{j\phi_{1}},\ldots,e^{j\phi_{N}})$. Thus, in this paper the \gls{ris} is employed for  channel-aware beamforming through the environment.
 
Before the data transmission phase starts, it is necessary to estimate the channels $\bH$ and $\bG$, and to configure the optimized phase shifts at the \gls{ris}. More details on channel estimation and \gls{ris} phase shifts configuration are provided in Section \ref{Sec:Overhead}. Nevertheless, at this stage it is important to stress that both channel estimation and resource optimization can be performed either at the transmitter or at the receiver, but not at the \gls{ris}. On the other hand, the \gls{ris} is interfaced with the transmitter through a controller with minimal signal processing, transmission/reception, and power storage capabilities.	The transmission/reception capabilities are needed in order to receive the configuration signals from the transmitter. The signal processing capabilities are needed in order to decode the configuration signals and configure the phase shifts of the \gls{ris}. The power storage capabilities are needed in order to operate the electronic circuits (switches or varactors) that make the surface reconfigurable. The controller is a key element to ensure the dynamic reconfigurability of the \gls{ris}, as a function of the propagation channel \cite[Figure 4]{RIS_GE_JSAC}. However, feeding back the optimized phase matrix $\bPhi$ to the RIS before the data transmission phase, may introduce a non-negligible overhead to the communication phase, especially for large $N$. Let us denote by $T_{F}$ the duration of the feedback phase, which depends on the power $p_{F}$ used during the feedback phase and on the bandwidth $B_{F}$ of the feedback channel. Moreover, let us denote by $T_{E}$ the duration of the channel estimation phase prior to feedback and communication. Mathematical expressions of $T_{F}$ and $T_{E}$ are provided in Section \ref{Sec:Overhead}. Then, denoting by $T$ the total duration of the time slot comprising channel estimation, feedback, and data communication, the system achievable rate and energy efficiency are expressed as 
\begin{align}\label{Eq:BitsTx}
&R(p,\!B,\!p_{F},\!B_{F},\!\bPhi,\!\bq,\!\bw)=\notag\\
&\left(1-\frac{T_{E}+T_{F}}{T}\!\right)B\log\left(1+\frac{p|\bw^{H}\bG\bPhi\bH\bq|^{2}}{B N_{0}}\!\right)\\
\label{Eq:EEObj}
&\text{EE}(p,B,\!p_{F},\!B_{F},\!\bPhi,\!\bq,\!\bw)=\frac{R(p,B,p_{F},B_{F},\bPhi,\bq,\bw)}{P_{tot}(p,B,p_{F},B_{F})},
\end{align}
wherein $P_{tot}$ denotes the total power consumption in the whole timeframe $T$, which is equal to
\begin{align}\label{Eq:Etot}
&P_{tot}(p,\!B,\!p_{F},\!B_{F})\!=\!P_{E}\!+\!\frac{(T\!-\!T_{E}\!-\!T_{F})}{T}\mu p\!+\!\frac{\mu_{F}p_{F}T_{F}}{T}\!+\!P_{c}\;,
\end{align}
since a power $p$ is used for $T-T_{E}-T_{F}$ seconds, with transmit amplifier efficiency $1/\mu$, a power $p_{F}$ is used for $T_{F}$ seconds, with transmit amplifier efficiency $1/\mu_{F}$, while a hardware static power $P_{c}$ is consumed for the whole interval $T$, and $P_{E}$ accounts for the energy consumption for channel estimation, which is further detailed in Section \ref{Sec:Overhead}. This work optimizes the transmit and feedback powers and bandwidths $p,p_{F},B,B_{F}$, the \gls{ris} matrix $\bPhi$, and the precoding and decoding vectors $\bq,\bw$, in order to maximize the rate \eqref{Eq:BitsTx}, the energy efficiency \eqref{Eq:EEObj}, and derive the rate-energy Pareto-region.  

\section{Optimization of $\bPhi$, $\bq$, $\bw$}\label{Sec:PhaseOpt}
As a first step, let us fix $p,p_{F},B,B_{F}$, and focus on optimizing the RIS phase matrix $\bPhi$, the unit-norm beamforming vector $\bq$, and the unit-norm decoding vector $\bw$. 
Since $\bPhi,\bq,\bw$ do not appear in the denominator of the energy efficiency, but only in the numerator, which coincides with the system  rate, both rate and energy efficiency maximization are cast as 
\beq\label{Prob:PhaseOpt}
\ds\max_{\small(\bPhi,\bq,\bw):\|\bq\|=\|\bw\|=1,\,\phi_{n}\in[0,2\pi],\,\forall n}\;|\bw^{H}\bG\bPhi\bH\bq|^{2}
\eeq
Denoting by $\lambda_{A,max}$ the largest singular value of $\bA$, it holds 
\begin{align}\label{Eq:Aopt}
\max_{\small(\bw,\bq)\;:\;\|\bw\|=\|\bq\|=1}|\bw^{H}\bA\bq|^{2}&\leq \max_{\small(\bw,\bq)\;:\;\|\bw\|=\|\bq\|=1}\|\bw\|^{2}\|\bA\bq\|^{2}\notag\\
&\!\!\!\!\!\!\leq \max_{\small\bq\;:\;\|\bq\|=1}\|\bA\bq\|^{2}=\lambda_{A,max}^{2},
\end{align}
where we have used Cauchy-Schwarz inequality, the constraint that $\|\bw\|=1$, and the fact that the maximum of $\|\bA\bq\|$ with respect to the set of unit-norm vectors $\bq$ is the spectral norm of  $\bA$, i.e. the largest singular value of $\bA$, \cite[pag. 148]{Horn1991}. Then, for any $\bA$, the optimal $\bq$ and $\bw$ are the dominant right and left eigenvector of $\bA$, since this achieves the upper-bound in \eqref{Eq:Aopt}.  However, optimally maximizing the largest singular value of $\bA=\bG\bPhi\bH$ with respect to $\bPhi$ appears prohibitive. Moreover, this would not yield any closed-form expression for $\bPhi$, $\bq$, $\bw$, which hinders the analytical evaluation of the ultimate performance of \gls{ris}-based networks. Thus, we propose two closed-form approaches for optimizing an upper-bound or a lower-bound of the objective of  \eqref{Prob:PhaseOpt}. 
\subsection{Optimizing an upper-bound of the objective of \eqref{Prob:PhaseOpt}}\label{Sec:UpperPhase}
Let $\bH=\sum_{j=1}^{r_{H}}\mu_{j,H}\bu_{j,H}\bv_{j,H}^{H}$, $\bG=\sum_{i=1}^{r_{G}}\mu_{i,G}\bu_{i,G}\bv_{i,G}^{H}$ be the \glspl{svd} of $\bH$ and $\bG$, with $r_{H}=\text{rank}(\bH)$,  $r_{G}=\text{rank}(\bG)$. Then, it holds that 
\begin{align}
&|\bw^{H}\bG\bPhi\bH\bq|^{2}=\left|\sum_{i=1}^{r_{G}}\sum_{j=1}^{r_{H}}\mu_{i,G}\mu_{j,H}\bw^{H}\bu_{i,G}\bv_{i,G}^{H}\bPhi\bu_{j,H}\bv_{j,H}^{H}\bq\right|^{2}\notag\\
&\stackrel{(a)}{\leq}\left(\sum_{i=1}^{r_{G}}\sum_{j=1}^{r_{H}}\!\mu_{i,G}\mu_{j,H}\!\left|\bw^{H}\bu_{i,G}\right|\!\left|\bv_{i,G}^{H}\bPhi\bu_{j,H}\right|\!\left|\bv_{j,H}^{H}\bq\right|\!\!\right)^{2}\notag\\
&\!\stackrel{(b)}{\leq}\!\!r_{G}r_{H}\!\!\sum_{i=1}^{r_{G}}\!\sum_{j=1}^{r_{H}}\!\mu_{i,G}^{2}\!\mu_{j,H}^{2}\!\left|\bw^{H}\bu_{i,G}\!\right|^{2}\!\left|\bv_{i,G}^{H}\bPhi\bu_{j,H}\!\right|^{2}\!\left|\bv_{j,H}^{H}\bq\right|^{2}\label{Eq:PhaseUpper}
\end{align}
wherein Inequality $(a)$ is due to the triangle inequality, while Inequality $(b)$ is a special case of Cauchy-Schwarz inequality.\footnote{Cauchy-Schwarz inequality states that $(\sum_{m=1}^{M}a_{m}b_{m})^{2}\leq(\sum_{m=1}^{M}a_{m}^{2})(\sum_{m=1}^{M}b_{m}^{2})$, for any non-negative numbers $\{a_{m},b_{m}\}_{m=1}^{M}$.  Then, by taking $b_{m}=1$ for all $m$, we obtain $(\sum_{m=1}^{M}a_{m})^{2}\leq M\sum_{m=1}^{M}a_{m}^{2}$} In the following, we derive a closed-form solution for the maximization of the bound in \eqref{Eq:PhaseUpper} with respect to $\bPhi$, $\bw$, $\bq$. We start with the following lemma.
\begin{lemma}\label{Lem:Phases}
Consider $c_{j}\geq 0$ and $x_{j}\geq 0$ for all $j=1,\ldots,J$, with $\sum_{j=1}^{J}x_{j}\leq 1$. Then it holds that $\max\;\sum_{j=1}^{J}c_{j}x_{j}\leq c_{\bar{j}}$, with $\bar{j}$ such that $c_{\bar{j}}\geq c_{j}$ for all $j=1,\ldots,J$.
\end{lemma}
\begin{IEEEproof}
Since $c_{\bar{j}}\geq c_{j}$ for all $j=1,\ldots,J$, there exist non-negative $\epsilon_{1},\ldots \epsilon_{J}$ such that $c_{j}=c_{\bar{j}}-\epsilon_{j}$, for all $j=1,\ldots,J$. Then, the result is shown as follows
\begin{align}
\sum_{j=1}^{J}c_{j}x_{j}&\!=\!c_{\bar{j}}x_{\bar{j}}\!+\!\sum_{j\neq\bar{j}}^{J}(c_{\bar{j}}-\epsilon_{j}) x_{j}
\!=\!c_{\bar{j}}\sum_{j=1}^{J}x_{j}\!-\!\sum_{j=2}^{J}\epsilon_{j}x_{j}\leq c_{\bar{j}}\;.\notag
\end{align}
\end{IEEEproof}
The optimal $\bPhi$, $\bq$, $\bw$ for the upper-bound in \eqref{Eq:PhaseUpper} are as follows.
\begin{proposition}\label{Prop:PahsesMIMO}
For any $p,B,p_{F},B_{F}$, defining
\begin{align}
\bar{j}(i)&=\text{argmax}_{j}\mu_{j,H}^{2}\!\left(\sum_{n=1}^{N}\left|\bv_{i,G}^{(n)}\right|\left|\bu_{j,H}^{(n)}\right|\right)^{2}\!\!\!\!,\!\forall i\!=\!1,\ldots,r_{G}\\
&\bar{i}=\text{argmax}_{i}\;\mu_{i,G}^{2}\mu_{\bar{j}(i),H}^{2}\left(\sum_{n=1}^{N}\left|\bv_{i,G}^{(n)}\right|\left|\bu_{\bar{j}(i),H}^{(n)}\right|\right)^{2}
\end{align}
the global maximizer of the upper-bound in \eqref{Eq:PhaseUpper} is obtained by setting $\bq=\bv_{\bar{j}(\bar{i}),H}$, $\bw=\bu_{\bar{i},G}$, and $\phi_{n}=-\angle{\left\{\bv_{\bar{i},G}^{*(n)}\bu_{\bar{j}(\bar{i}),H}^{(n)}\right\}}$, with $^{(*)}$ denoting complex conjugate.
\end{proposition}
\begin{IEEEproof}
Neglecting the inessential factors $r_{G}r_{H}$, we observe that
\begin{align}\label{Eq:PhasesMIMO}
&\sum_{i=1}^{r_{G}}\sum_{j=1}^{r_{H}}\mu_{i,G}^{2}\mu_{j,H}^{2}\!\left|\bw^{H}\bu_{i,G}\right|^{2}\!\left|\bv_{i,G}^{H}\bPhi\bu_{j,H}\right|^{2}\!\left|\bv_{j,H}^{H}\bq\right|^{2}\\
&\leq\!\sum_{i=1}^{r_{G}}\!\mu_{i,G}^{2}|\bw^{H}\bu_{i,G}|^{2}\!\underbrace{\sum_{j=1}^{r_{H}}\!\mu_{j,H}^{2}\!\max_{\bPhi}\!\left\{\!|\bv_{i,G}^{H}\bPhi\bu_{j,H}|^{2}\!\right\}\!|\bv_{j,H}^{H}\bq|^{2}}_{y_{i}}\notag
\end{align}
wherein the inequality follows upon taking the maximum over $\bPhi$. Next, for any $i=1,\ldots,r_{G}$, the term $y_{i}$ defined in the last line of \eqref{Eq:PhasesMIMO} can be upper-bounded as
\begin{align}\label{Eq:PhasesMIMO2}
y_{i}&=\sum_{j=1}^{r_{H}}\mu_{j,H}^{2}\max_{\bPhi}\left\{|\bv_{i,G}^{H}\bPhi\bu_{j,H}|^{2}\right\}|\bv_{j,H}^{H}\bq|^{2}\notag\\
&\stackrel{(a)}{\leq}\mu_{\bar{j}(i),H}^{2}\max_{\bPhi}\left\{|\bv_{i,G}^{H}\bPhi\bu_{\bar{j}(i),H}|^{2}\right\}\notag\\&\stackrel{(b)}{=}\mu_{\bar{j}(i),H}^{2}\left(\sum_{n=1}^{N}\left|\bv_{i,G}^{(n)}\right|\left|\bu_{\bar{j}(i),H}^{(n)}\right|\right)^{2}\;,
\end{align}
wherein (b) follows because, for any $i=1,\ldots,r_{G}$, the optimal $\bPhi$ is the one that compensates the phase shifts between the components of $\bv_{i,G}^{H}$ and of $\bu_{\bar{j}(i)}$, while (a) follows from Lemma \ref{Lem:Phases} because $\sum_{j=1}^{r_{H}}|\bv_{j,H}^{H}\bq|^{2}\leq\|\bq\|^{2}=1$, since, for all $j=1,\ldots,r_{H}$, $\bv_{j,H}^{H}\bq$ is the projection of the unit-norm vector $\bq$ onto the unit-norm vector $\bv_{j,H}$. Plugging \eqref{Eq:PhasesMIMO2} into \eqref{Eq:PhasesMIMO}, yields 
\begin{align}
|\bw^{H}\bG\bPhi\bH\bq|^{2}\!&\leq\!\sum_{i=1}^{r_{G}}\mu_{i,G}^{2}\mu_{\bar{j}(i),H}^{2}\!\!\left(\sum_{n=1}^{N}\!\left|\bv_{i,G}^{(n)}\!\right|\!\left|\bu_{\bar{j}(i),H}^{(n)}\!\right|\!\right)^{2}\!\!\!|\bw^{H}\bu_{i,G}|^{2}\notag\\
&\leq\mu_{\bar{i},G}^{2}\mu_{\bar{j}(\bar{i}),H}^{2}\!\left(\sum_{n=1}^{N}\left|\bv_{\bar{i},G}^{(n)}\right|\left|\bu_{\bar{j}(\bar{i}),H}^{(n)}\right|\!\right)^{2}
\end{align}
wherein the last inequality holds by Lemma \ref{Lem:Phases}. Finally, the result follows since all inequalities hold with equality upon choosing $\bPhi, \bq, \bw$ as in the thesis of the proposition.
\end{IEEEproof}

\subsection{Optimizing a lower-bound of the objective of  \eqref{Prob:PhaseOpt}}\label{Sec:LowerPhase}
Define $\bg_{w}=\bG^{H}\bw$, $\bh_{q}=\bH\bq$, and observe that, for any fixed $\bq$ and $\bw$, the optimal $\bPhi$ for Problem \eqref{Prob:PhaseOpt} is such that $\phi_{n}=-\angle{\{\bg_{w}^{*}(n)\bh_{q}(n)\}}$, for all $n=1,\ldots,N$. Next, denoting by $\bh_{n}^{T}\in\mathcal{R}^{1\times N}$ and $\bg_{n}\in\mathcal{R}^{N\times 1}$ the $n$-th row of $\bH$ and the $n$-th column of $\bG$, respectively, with $n=1,\ldots,N$, it holds that $\bg_{w}(n)=\bw^{H}\bg_{n}$ and $\bh_{q}(n)=\bh_{n}^{T}\bq$. Then, we obtain
\begin{align}
&\max_{\bq,\bw,\bPhi}\!|\bw^{H}\bG\bPhi\bH\bq|^{2}\!=\!\max_{\bq,\bw}\!\left(\!\max_{\bPhi}|\bg_{w}^{H}\bPhi\bh_{q}|^{2}\!\right)\!\!\stackrel{(a)}=\\
&\max_{\bq,\bw}\!\left(\!\sum_{n=1}^{N}|\bw^{H}\bg_{n}\bh_{n}^{T}\bq|\!\right)^{\!\!2}\!\!\!\stackrel{(b)}\geq\!\max_{\bq,\bw}\left|\bw^{H}\!\!\left(\!\sum_{n=1}^{N}\bg_{n}\bh_{n}^{T}\!\right)\!\!\bq\right|^{2}\notag
\end{align}
where $(a)$ follows by using the maximizer with respect to $\bPhi$, i.e. $\phi_{n}=-\angle{\{\bg_{w}^{*}(n)\bh_{q}(n)\}}$, and $(b)$ is due to the triangle inequality. Then, by similar steps as those that led to \eqref{Eq:Aopt}, the final maximization is obtained when $\bq,\bw$ are the dominant right and left eigenvector of $\sum_{n=1}^{N}\bg_{n}\bh_{n}^{T}$.

\subsection{Tackling \eqref{Prob:PhaseOpt} by alternating maximization}\label{Sec:AOPhase}
As a benchmark solution, let us maximize $|\bw^{H}\bG\bPhi\bH\bq|^{2}$ by alternatively optimizing $\bPhi$, for fixed $\bw,\bq$, and then $\bw,\bq$, for fixed $\bPhi$. For fixed $\bPhi$, the optimal $\bw$ and $\bq$ are derived as the dominant left and right eigenvectors of the matrix $\bA=\bG\bPhi\bH$, as shown in \eqref{Eq:Aopt}. Instead, for fixed $\bw$ and $\bq$, the problem amounts to maximizing $\bg_{w}^{H}\bPhi\bh_{q}$, which yields $\phi_{n}=-\angle{\{\bg_{w}^{*}(n)\bh_{q}(n)\}}$, for all $n$, as shown in Section \ref{Sec:LowerPhase}. Thus, alternating maximization leads to Algorithm \ref{Alg:OptPhase}.
\begin{algorithm}\caption{Alternating optimization of $\bPhi,\bq,\bw$}
\begin{algorithmic}\small\label{Alg:OptPhase}
\STATE \texttt{Initialize $\bw$ and $\bq$ to feasible values.}
\REPEAT
\STATE $\bg_{w}=\bG^{H}\bw$ and $\bh_{q}=\bH\bq$; \texttt{ set } $\phi_{n}=-\angle{\{\bg_{w}^{*}(n)\bh_{q}(n)\}}$ \texttt{ for all $n=1,\ldots,N$}; $\bA=\bG\bPhi\bH$;
\STATE \texttt{Set $\bw$ and $\bq$ as the left and right dominant eigenvectors of $\bA$}; 
\UNTIL{Convergence}
\end{algorithmic}
\end{algorithm}
\subsection{Overhead modeling}\label{Sec:Overhead}
This section derives a mathematical expression for $T_{F}$, $T_{E}$, $P_{E}$. Without loss of generality, we assume that channel estimation and resource optimization takes place at the transmitter.\footnote{A similar argument would hold in the case in which channel estimation and resource optimization took place at the receiver.}

As for $T_{F}$, after resource optimization, the transmitter sends a control signal to the RIS to configure the phase shifts.\footnote{Here we neglect the feedback of the receive filter $\bw$ to the receiver, because, first it is negligible with respect to the feedback of the RIS phase shifts, since typically $N>>N_{R}$, and, second, because the focus of this work is on the RIS and on evaluating the feedback required to operate it.} Denoting by $h_{F}$ the scalar feedback channel from the RIS to the transmitter, it holds that $T_{F}=\frac{Nb_{F}}{B_{F}\log\left(1+\frac{p_{F}|h_{F}|^{2}}{N_{0}B_{F}}\right)}$, with $b_{F}$ the number of feedback bits for each reflecting element of the \gls{ris} and $N_{0}$ the noise power spectral density. As anticipated, $T_{F}$ depends on $p_{F}$, $B_{F}$, which complicates the mathematical structure of the rate in \eqref{Eq:BitsTx} and the energy efficiency in \eqref{Eq:EEObj}, complicating the optimization of these two metrics, and of their trade-off, with respect to $p,p_{F},B,B_{F}$. The optimization of $p,p_{F},B,B_{F}$ for the maximization of the rate, the energy efficiency, and their trade-off is addressed in Section \ref{Sec:ContinuousRate}.

As for $T_{E}$, it is affected by the specific channel estimation protocol in use. As an example, we consider that the pilot tones are sent by the receiver to the transmitter, but a similar analysis applies to the case in wich the transmitter send pilot tones to the receiver. Moreover, we consider that, during the estimation phase, the RIS does not apply any phase shift, i.e. $\phi_{n}=0$ for all $n=1,\ldots,N$, \cite{He2019}. In the following, two different channel estimation protocols are considered: 
\begin{itemize}
\item [(a)] Let us consider the simple case in which the receiver sends pilot tones sequentially, one after the other, to the transmitter, through the RIS. Thus, the $NN_{T}N_{R}$ product channels $h_{nt,n}g_{n,nr}$ are estimated sequentially, with $h_{nt,n}$ denoting the channel from the $nt$-th transmit antenna to the $n$-th RIS elements, and $g_{n,nr}$ denoting the channel from the $n$-th RIS element to the $nr$-th receive antenna. Moreover, one additional pilot tone is required for the transmitter to estimate the feedback channel. Therefore, denoting by $T_{0}$ the duration of each pilot tone, it holds $T_{E}=(N_{T}NN_{R}+1)T_{0}$. It should be remarked that the knowledge of the product channels $h_{nt,n}g_{n,nr}$ is enough to reconstruct the matrix $\bA$ in \eqref{Eq:Aopt} and, therefore, to optimally solve \eqref{Prob:PhaseOpt} with respect to the phase shifts of the \gls{ris}, the beamforming vector, and the receive filter. Finally, the energy consumption for channel estimation can be modeled as $P_{E}=P_{0}(1+NN_{T}N_{R})T_{0}/T$, with $P_{0}$ the power of each pilot tone. In particular, the overhead needed for channel estimation is estimated based on the channel state information needed to optimally solve \eqref{Prob:PhaseOpt}. However, our algorithms work also with recently proposed channel estimation algorithms, e.g., \cite{He2019}. 
\item[(b)] The case in which the receiver transmits $N_{R}$ orthogonal pilots in parallel, which are jointly processed at the transmitter. Then, for all $nr=1,\ldots,N_{R}$, the pilot from the $nr$-th antenna of the receiver allows estimating the product channels $g_{n,nr}h_{nt,n}$, for all $n=1,\ldots,N$ and $nt=1,\ldots,N_{T}$. Thus, in this case it holds $T_{E}=(N+1)T_{0}$, since all pilots are transmitted at the same time. On the other hand, $P_{E}=(NN_{R}+1)P_{0}T_{0}/T$, because the $N_{R}$ pilots are transmitted at the same time, each with power $NP_{0}$.
\end{itemize}
Thus, based on the expressions of $T_{F}$ and $T_{E}$, the power consumption in \eqref{Eq:Etot} becomes 
\beq
P_{tot}=P_{E}+\frac{b_{F}N(\mu_{F}p_{F}-\mu p)}{TB_{F}\log\left(1+\frac{p_{F}|h_{F}|^{2}}{B_{F}N_{0}}\right)}+\mu p\left(1-\frac{T_{E}}{T}\right)+P_{c}\;.\notag
\eeq 
\section{Optimization of $p,p_{F},B,B_{F}$.}\label{Sec:ContinuousRate}
After optimizing $\bPhi,\bq,\bw$ by any of the methods developed in Section \ref{Sec:PhaseOpt}, we are left with the problem of optimizing the transmit powers $p,p_{F}$, and the bandwidths $B,B_{F}$. It should be stressed that, as already mentioned, the optimized $\bPhi,\bq,\bw$ that are obtained from any of the algorithms developed in Section \ref{Sec:PhaseOpt}, do not depend on any of the variables $p,p_{F},B,B_{F}$, but only on the channels $\bH$ and $\bG$. Moreover, as already mentioned, the optimized $\bPhi,\bq,\bw$ are the same for both the rate and the energy efficiency. Thus, it is possible to simply plug in the optimized $\bPhi,\bq,\bw$ into the objective to maximize, thus effectively decoupling the optimization of $p,p_{F},B,B_{F}$ from the optimization of $\bPhi,\bq,\bw$. On the other hand, unlike $\bPhi,\bq,\bw$, the optimization of $p,p_{F},B,B_{F}$ depends on whether the goal is optimizing the rate, the energy efficiency, or their trade-off. Therefore, these problems are treated separately. 
\vspace{-0.5cm}
\subsection{Rate maximization}\label{Sec:RateOpt}
The rate maximization problem is stated as the following optimization program
\begin{subequations}\label{Eq:ProbContinuous2}
\begin{align}
&\ds\max_{p,B,p_{F},B_{F}} R(p,B,p_{F},B_{F},\bPhi_{\text{opt}},\bq_{\text{opt}},\bw_{\text{opt}})\label{Eq:ProbContinuous2a}\\
&\;\text{s.t.}\;p+p_{F}\leq P_{max}\;,B+B_{F}\leq B_{max}\label{Eq:ProbContinuous2b}\\
&\quad\;\;p\geq 0\;,\;p_{F}\geq0\;, B\geq 0\;,\;B_{F}\geq 0\label{Eq:ProbContinuous2c}\\
&\quad\;\;\frac{b_{F}N}{TB_{F}\log\left(1+\frac{p_{F}|h_{F}|^{2}}{B_{F}N_{0}}\right)}\leq 1-\frac{T_{E}}{T}\label{Eq:ProbContinuous2f}\;,
\end{align}
\end{subequations}
Expressing \eqref{Eq:ProbContinuous2a} as a function of the optimization variables $p,p_{F},B,B_{F}$ yields
\begin{align}\label{Eq:Rateobjective}
R(p,B,p_{F},B_{F})&=\left(\beta-\frac{d}{B_{F}\log\left(1+\frac{p_{F}|h_{F}|^{2}}{N_{0}B_{F}}\right)}\right)
\times\notag\\
&B\log\!\left(1+\!\frac{p|\bw_{\text{opt}}^{H}\bG\bPhi_{\text{opt}}\bH\bq_{\text{opt}}|^{2}}{B N_{0}}\right)\;,
\end{align}
wherein $\beta=1-T_{E}/T$ and $d=b_{F}N/T$. Thus, being the product of two functions, the objective of  \eqref{Eq:ProbContinuous2} is not jointly concave in all optimization variables, which makes Problem \eqref{Eq:ProbContinuous2} challenging to solve with affordable complexity. Indeed, the product of functions is in general not concave even in the simple case in which the individual factors are concave. Moreover, in the case at hand, the concavity of the two factors defining \eqref{Eq:Rateobjective} is not clear, either. Thus, in order to solve \eqref{Eq:ProbContinuous2}, it is not possible to directly use standard convex optimization algorithms. In the rest of this section we show that it is possible to reformulate Problem \eqref{Eq:ProbContinuous2} into a convex optimization problem without any loss of optimality. To this end, some preliminary lemmas are needed.
\begin{lemma}\label{Prop:G1}
The function $R(p,B,p_{F},B_{F})$ is jointly increasing and jointly concave in $(p,B)$.
\end{lemma}
\begin{IEEEproof}
Neglecting inessential constant terms (with respect to $p$ and $B$), and defining
\beq\label{Eq:c}
c=\ds\frac{|\bw_{\text{opt}}^{H}\bG\bPhi_{\text{opt}}\bH\bq_{\text{opt}}|^{2}}{N_{0}}\;,
\eeq
Eq. \eqref{Eq:ProbContinuous2a} is equivalent to the function $g_{1}(p,B)=B\log\left(1+\frac{pc}{B}\right)$, which is the perspective of the concave function $\log\left(1+pc\right)$ \cite{boyd2004convex}. Thus, since the perspective operator preserves concavity, $g_{1}$ is jointly concave in $(p,B)$. Moreover, $g_{1}$ is clearly increasing in $p$, while inspecting the derivative of $g_{1}$ with respect to $B$, and exploiting that $(1+y)\log(1+y)\geq y$ for any $y\geq 0$, shows that $g_{1}$ is increasing in $B$.
\end{IEEEproof}

\begin{lemma}\label{Prop:G2}
The function $R(p,B,p_{F},B_{F})$ is jointly increasing and jointly concave in $(p_{F},B_{F})$.
\end{lemma}
\begin{IEEEproof}
Neglecting inessential constant terms (with respect to $p_{F}$ and $B_{F}$), it can be seen that, upon defining $a=|h_{F}|^{2}/N_{0}$, the function in \eqref{Eq:ProbContinuous2a} is equivalent to 
\beq\label{Eq:F2}
\beta-\frac{d}{B_{F}\log\left(1+a\frac{p_{F}}{B_{F}}\right)}\;.
\eeq
Showing the joint concavity of \eqref{Eq:F2} with respect to $(p_{F},B_{F})$ is equivalent to showing that the function $g_{2}(p_{F},B_{F})=\frac{1}{B_{F}\log\left(1+a\frac{p_{F}}{B_{F}}\right)}=\frac{1}{z(p_{F},B_{F})}$, is jointly convex in $(p_{F},B_{F})$. After some elaborations, the Hessian matrix of $g_{2}$ is written as given in \eqref{Eq:Hessian}, shown at the top of the next page,
\begin{figure*}
\small
\begin{align}\label{Eq:Hessian}
&{\cal H}=\frac{1}{(B_{F}+ap_{F})^{2}z^{3}(p_{F},B_{F})}\times\\
&\left[\begin{array}{cc}a^{2}B_{F}z(p_{F},B_{F})+2a^{2}B_{F}^{2} & -a^{2}p_{F}z(p_{F},B_{F})+2aB_{F}(B_{F}+ap_{F})z_{B_{F}}^{'}(p_{F},B_{F})\\
 -a^{2}p_{F}z(p_{F},B_{F})+2aB_{F}(B_{F}+ap_{F})z_{B_{F}}^{'}(p_{F},B_{F}) & \frac{a^{2}p_{F}^{2}}{B_{F}}z(p_{F},B_{F})+2(B_{F}+ap_{F})^{2}\left(z_{B_{F}}^{'}(p_{F},B_{F})\right)^{2}
\end{array}\right]\notag
\end{align}
\end{figure*}
wherein $z_{B_{F}}^{'}(p_{F},B_{F})=\log\left(1+a\frac{p_{F}}{B_{F}}\right)-\frac{ap_{F}}{B_{F}+ap_{F}}$ is the first-order derivative of $z$ with respect to $B_{F}$. Clearly, the entry $(1,1)$ of ${\cal H}$ is non-negative. Thus, ${\cal H}$ is positive semi-definite if its determinant is non-negative. Then, since the second derivative of $z$ with respect to $B_{F}$ can be written as
\beq
z_{B_{F}}^{''}(p_{F},B_{F})=-\frac{ap_{F}^{2}}{B_{F}(B_{F}+ap_{F})^{2}}\;,
\eeq
after some elaborations, enforcing that the Hessian of ${\cal H}$ is non-negative leads to the condition $(B_{F}+ap_{F})^{2}(z^{'}(p_{F},B_{F}))^{2}+2a^{2}p_{F}^{2}+2ap_{F}z^{'}(p_{F},B_{F})\geq 0$. This holds if $z^{'}(p_{F},B_{F})\geq 0$, which is true by virtue of the inequality $(1+y)\log(1+y)\geq y$. Moreover,  $z^{'}(p_{F},B_{F})\geq 0$ implies that $z(p_{F},B_{F})$ is increasing in $B_{F}$, while it is clearly increasing in $p_{F}$.
\end{IEEEproof}

Leveraging Lemmas \ref{Prop:G1} and \ref{Prop:G2}, it is possible to equivalently reformulate Problem \eqref{Eq:ProbContinuous2} into a convex problem, which can then be efficiently solved by means of any convex optimization method. To this end, the first step is to observe that taking the logarithm of the objective in \eqref{Eq:ProbContinuous2a} does not change the optimal solutions of \eqref{Eq:ProbContinuous2a}, since the logarithm is an increasing function. Then, an equivalent reformulation of \eqref{Eq:ProbContinuous2a} is the following problem
\begin{subequations}\label{Eq:ProbContinuous3}
\begin{align}
&\ds\max_{p,B,p_{F},B_{F}} \log(\beta-d g_{2}(p_{F},B_{F}))+\log(g_{1}(p,B))\label{Eq:ProbContinuous3a}\\
&\;\text{s.t.}\;p+p_{F}\leq P_{max}; B+B_{F}\leq B_{max}\label{Eq:ProbContinuous3b}\\
&\quad\;\;p\geq 0\;,\;p_{F}\geq0; B\geq 0\;,\;B_{F}\geq 0\label{Eq:ProbContinuous3c}\\
&\quad\;\;\frac{d}{B_{F}\log\left(1+\frac{p_{F}|h_{F}|^{2}}{B_{F}N_{0}}\right)}\leq \beta\;,\label{Eq:ProbContinuous3d}
\end{align}
\end{subequations}
which is a convex optimization problem by virtue of Lemmas \ref{Prop:G1} and \ref{Prop:G2}. Indeed, \eqref{Eq:ProbContinuous3a} is a concave function since Lemmas \ref{Prop:G1} and \ref{Prop:G2} ensure that both summands are concave. Also, all the constraints in \eqref{Eq:ProbContinuous3b}-\eqref{Eq:ProbContinuous3c} are linear, while \eqref{Eq:ProbContinuous3d} is convex thanks to Lemma \ref{Prop:G2}. Thus, Problem \eqref{Eq:ProbContinuous3} is a convex problem with the same set of solutions as Problem \eqref{Eq:ProbContinuous2}, but the advantage that it can be solved by convex optimization theory.

Finally, in order to further simplify the solution of \eqref{Eq:ProbContinuous3}, we observe that the optimal solution of \eqref{Eq:ProbContinuous3} is such that \eqref{Eq:ProbContinuous3b} and \eqref{Eq:ProbContinuous3c} must be fulfilled with equality, since the objective function is increasing in all arguments and \eqref{Eq:ProbContinuous3d} is decreasing in both $B_{F}$ and $p_{F}$. Thus, Problem \eqref{Eq:ProbContinuous3} can be reformulated, without loss of optimality, as
\begin{subequations}\label{Eq:ProbContinuous4}
\begin{align}
&\ds\max_{p,B} \log(\beta-d g_{2}(P_{max}-p,B_{max}-B))+\log(g_{1}(p,B))\label{Eq:ProbContinuous4a}\\
&\;\text{s.t.}\;0\leq p\leq P_{max}\;,\; 0\leq B\leq B_{max}\label{Eq:ProbContinuous4b}\\
&\;\quad\;\;\frac{d}{(B_{max}-B)\log\left(1+\frac{(P_{max}-p)|h_{F}|^{2}}{(B_{max}-B)N_{0}}\right)}\leq \beta\;,
\end{align}
\end{subequations}
which has only two optimization variables. Upon solving \eqref{Eq:ProbContinuous4}, the optimal feedback power and bandwidth are retrieved as $p_{F}=P_{max}-p$ and $B_{F}=B_{max}-B$. Problem \eqref{Eq:ProbContinuous4} is clearly still a convex problem, since it is obtained from the convex Problem \eqref{Eq:ProbContinuous3} upon applying the linear variable transformations $p_{F}=P_{max}-p$ and $B_{F}=B_{max}-B$, and linear transformations are well-known to preserve convexity. Finally, after developing a method for solving \eqref{Eq:ProbContinuous4} with affordable complexity, in the last part of this section we focus on obtaining closed-form solutions for the special cases of Problem \eqref{Eq:ProbContinuous4} obtained by considering the optimization of the transmit powers for fixed bandwidths and vice-versa. Closed-form solutions can be obtained as follows.

\subsubsection{Optimization for fixed $B$ and $B_{F}$}
Fixing $B$ and $B_{F}$, Problem \eqref{Eq:ProbContinuous4} reduces to
\begin{subequations}\label{Eq:Power}
\begin{align}
&\ds\max_{p}\; \log(\beta-d g_{2}(P_{max}-p,B_{F}))+\log(g_{1}(p,B))\label{Eq:aPower}\\
&\;\text{s.t.}\;0\leq p\leq P_{max}-\frac{B_{F}N_{0}}{|h_{F}|^{2}}\left(e^{\frac{d}{B_{F}\beta}}-1\right)\;.\label{Eq:bPower}
\end{align}
\end{subequations}
\begin{proposition}\label{Prop:PowerOpt}
Let $\bar{p}$ be the unique stationary point of \eqref{Eq:aPower}. Then, Problem \eqref{Eq:Power} has a unique solution given by $p^{*}=\min(\bar{p},P_{max}-p_{min})$, with $\bar{p}$ the unique solution of Eq. \eqref{Eq:StationaryCondP}, shown at the top of this page.
\begin{figure*}
\begin{equation}\label{Eq:StationaryCondP}
\frac{da}{\left(B_{F}\!+\!a(P_{max}\!-\!p)\right)\!\left(\!\beta B_{F}\log\!\left(\!1\!+\!a\frac{(P_{max}\!-\!p)}{B_{F}}\!\right)\!-\!d\right)\!\left(\!\log(1\!+\!(P_{max}\!-\!p)\frac{a}{B_{F}})\right)}\!=\!\frac{c}{(B\!+\!pc)\log\!\left(\!1\!+\!\frac{pc}{B}\!\right)}\;.
\end{equation}
\end{figure*}
\end{proposition}
\begin{IEEEproof}
Equating the first-order derivative of \eqref{Eq:aPower} to zero yields \eqref{Eq:StationaryCondP}, which has always a solution, since the left-hand-side is decreasing in $p$ and tending to $\infty$ for $p\to 0^{+}$, while the right-hand-side is increasing in $p$, being finite at $p=0$ and tending to $\infty$ for $p\to P_{max}$. Then, \eqref{Eq:aPower} has a unique solution $\bar{p}$, since \eqref{Eq:aPower} is a strictly concave function in $p$, as it is the sum of concave functions and $\log(g_{1}(p,B))$ is strictly concave in $p$. Finally, \eqref{Eq:StationaryCondP} shows that \eqref{Eq:aPower} is strictly increasing for $p<\bar{p}$ and strictly decreasing for $p>\bar{p}$. Thus, we can conclude that the unique solution of Problem \eqref{Eq:Power} is either $\bar{p}$, if $\bar{p}\leq P_{max}$, or it is $P_{max}$ itself.
\end{IEEEproof}
Finally, it holds $p_{F}^{*}=P_{max}-p^{*}$.

\subsubsection{Optimization for fixed $p$ and $p_{F}$}
Fixing $p$ and $p_{F}$, Problem \eqref{Eq:ProbContinuous4} reduces to
\begin{subequations}\label{Eq:Band}
\begin{align}
&\ds\max_{B,B_{F}}\; \log(\beta-dg_{2}(B_{max}-B,p_{F}))+\log(g_{1}(p,B))\label{Eq:aBand}\\
&\;\text{s.t.}\;0\leq B\leq B_{max}-\widehat{B}\;,\label{Eq:bBand}
\end{align}
\end{subequations}
with $\widehat{B}$ the unique\footnote{The uniqueness holds because the function at the left-hand-side is strictly decreasing, as it immediately follows from previous results.} value of $B$ that fulfills the following inequality with equality
\beq
(B_{max}-B)\log\left(1+\frac{p_{F}|h_{F}|^{2}}{N_{0}(B_{max}-B)}\right)\geq \frac{d}{\beta}\;.
\eeq
\begin{proposition}\label{Prop:BandOpt}
Problem \eqref{Eq:Band} has a unique solution given by $B^{*}=\min(\bar{B},B_{max}-\widehat{B})$, with $\bar{B}$ the unique stationary point of \eqref{Eq:aBand}.
\end{proposition}
\begin{IEEEproof}
The proof is similar to Proposition \ref{Prop:PowerOpt}. The objective  \eqref{Eq:aBand} is strictly concave, has a unique stationary point $\bar{B}$ given by the solution of the stationarity condition in Eq. \eqref{Eq:StationaryCondB}, shown at the top of next page, and is strictly increasing for $B<\bar{B}$ and strictly decreasing for $\bar{B}>\bar{B}$. 
\begin{figure*}
\beq\label{Eq:StationaryCondB}
\frac{\log\left(1+\frac{cp}{B}\right)-\frac{cp}{B+cp}}{B\log\left(1+\frac{cp}{B}\right)}=
\frac{\left(\log\left(1+\frac{ap_{F}}{B_{max}-B}\right)-\frac{ap_{F}}{B_{max}-B+ap_{F}}\right)d}{\left(\beta\left(B_{max}-B\right)\log\left(1+\frac{ap_{F}}{B_{max}-B}\right)-d\right)\left(B_{max}-B\right)\log\left(1+\frac{ap_{F}}{B_{max}-B}\right)}\;,
\eeq
\end{figure*}
\end{IEEEproof}
Finally, it holds $B_{F}^{*}=B_{max}-B^{*}$.

\subsection{Energy efficiency optimization}\label{Sec:EEOptimization}
Plugging again any of the allocations of $\bPhi,\bq,\bw$ developed in Section \ref{Sec:PhaseOpt} into the energy efficiency function, leads us to the following problem to solve
\begin{subequations}\label{Eq:EEcont}
\begin{align}
&\ds\max_{p,B,p_{F},B_{F}} \frac{R(p,\!B,\!p_{F},\!B_{F},\!\bPhi^{\text{opt}},\!\bq^{\text{opt}},\!\bw^{\text{opt}})}{P_{tot}(p,\!p_{F},\!B_{F})}\label{Eq:aEEcont}\\
&\;\text{s.t.}\;p+p_{F}\leq P_{max}\;,B+B_{F}\leq B_{max}\label{Eq:bEEcont}\\
&\quad\;\;p\geq 0\;,\;p_{F}\geq0\;,B\geq 0\;,\;B_{F}\geq 0\label{Eq:cEEcont}\\
&\quad\;\;\frac{d}{B_{F}\log\left(1+\frac{p_{F}|h_{F}|^{2}}{B_{F}N_{0}}\right)}\leq \beta\;.
\end{align}
\end{subequations}
It should be stressed that, in order to solve \eqref{Eq:EEcont}, it is not possible to employ the same approach used for rate maximization, because the presence of the denominator makes the logarithm of \eqref{Eq:aEEcont} not jointly concave in all optimization variables. Moreover, standard fractional programming algorithms are not directly applicable since they have limited complexity only when the numerator and the denominator of the objective to maximize are concave and convex functions, respectively. Unfortunately, in \eqref{Eq:EEcont}, neither the concavity of the numerator, nor the convexity of the denominator hold. Finally, a third issue that makes \eqref{Eq:EEcont} more challenging than the rate optimization problem is that, unlike the rate function, \eqref{Eq:aEEcont} is not monotonically increasing in either $p$ or $p_{F}$, and so it can not be guaranteed that, at the optimum, it holds $p+p_{F}=P_{max}$. On the other hand, \eqref{Eq:aEEcont} is increasing in $B$ and $B_{F}$, since, as shown in Section \ref{Sec:ContinuousRate}, the numerator is increasing in $B$ and $B_{F}$, while the denominator depends only on $B_{F}$ and decreases with $B_{F}$. Thus, at the optimum $B+B_{F}=B_{max}$ holds. Exploiting this and defining 
\beq\label{Eq:yAux}
y=(B_{max}-B)\log\left(1+\frac{p_{F}|h_{F}|^{2}}{(B_{max}-B)N_{0}}\right)\;,
\eeq
\eqref{Eq:EEcont} can be cast as
\begin{subequations}\label{Eq:ContinuousEE}
\begin{align}
&\ds\max_{p,B,p_{F},y} \frac{(\beta-\frac{d}{y})B\log\left(1+\frac{pc}{B}\right)}{\beta\mu p+P_{c}+\frac{d}{y}(\mu_{F}p_{F}-\mu p)}
\label{Eq:ContinuousEEa}\\
&\;\text{s.t.}\;p+p_{F}\leq P_{max}\label{Eq:ContinuousEEb}\\
&\quad\;\;0\leq B\leq B_{max}\;,p\geq 0\;,\;p_{F}\geq0\label{Eq:ContinuousEEc}\\
&\quad\;\;y=(B_{max}-B)\log\left(1+\frac{p_{F}|h_{F}|^{2}}{(B_{max}-B)N_{0}}\right)\;,\;y\geq \frac{d}{\beta} \label{Eq:ContinuousEEe}
\end{align}
\end{subequations}
wherein $P_{c}=NP_{c,n}+P_{c,0}+P_{E}$, and $c$ is given in \eqref{Eq:c}. Next, we also consider a relaxed version of \eqref{Eq:ContinuousEE} in which \eqref{Eq:ContinuousEEe} is reformulated into an inequality constraint, namely
\begin{subequations}\label{Eq:ContinuousEE2}
\begin{align}
&\ds\max_{p,B,p_{F},y} \frac{(\beta-\frac{d}{y})B\log\left(1+\frac{pc}{B}\right)}{\beta\mu p+P_{c}+\frac{d}{y}(\mu_{F}p_{F}-\mu p)}
\label{Eq:ContinuousEE2a}\\
&\;\text{s.t.}\;p+p_{F}\leq P_{max}\label{Eq:ContinuousEE2b}\\
&\quad\;\;0\leq B\leq B_{max}\;,p\geq 0\;,\;p_{F}\geq0\label{Eq:ContinuousEE2c}\\
&\quad\;\;y\leq (B_{max}-B)\log\left(1+\frac{p_{F}|h_{F}|^{2}}{(B_{max}-B)N_{0}}\right)\;,\;y\geq \frac{d}{\beta} \label{Eq:ContinuousEE2e}
\end{align}
\end{subequations}
which, unlike \eqref{Eq:ContinuousEE}, has a convex feasibility set, thanks to the fact that the first constraint in \eqref{Eq:ContinuousEEe} is an inequality constraint wherein the right-hand-side is a concave function. An important result is that, as shown in the coming proposition, \eqref{Eq:ContinuousEE} and \eqref{Eq:ContinuousEE2} are equivalent problems.
\begin{proposition}\label{Prop:EERelaxation}
Problem \eqref{Eq:ContinuousEE} and \eqref{Eq:ContinuousEE2} have the same set of optimal solutions.
\end{proposition}
\begin{IEEEproof}
The result follows by showing that any optimal solution of \eqref{Eq:ContinuousEE2} is such that
$y=(B_{max}-B)\log\left(1+\frac{p_{F}|h_{F}|^{2}}{(B_{max}-B)N_{0}}\right)$. To this end, let us  observe that \eqref{Eq:ContinuousEE2a} is monotonically increasing in $y$. Indeed, by dividing numerator and denominator by $(\beta-\frac{d}{y})$, \eqref{Eq:ContinuousEE2a} can be equivalently expressed as $\frac{B\log\left(1+\frac{pc}{B}\right)}{\mu p + \frac{P_{c}y}{\beta y-d}+\frac{d\mu_{F}p_{F}}{\beta y-d}}$,
which is strictly increasing in $y$. Based on this, the result follows proceeding by contradiction. Specifically, if $\bar{y}$ were a solution of \eqref{Eq:ContinuousEE2}, but $y<(B_{max}-B)\log\left(1+\frac{p_{F}|h_{F}|^{2}}{(B_{max}-B)N_{0}}\right)$, then it would be possible to find a feasible $y^{*}>\bar{y}$. Since \eqref{Eq:ContinuousEE2a} is increasing in $y$, $y^{*}$ would yield a larger objective value than $\bar{y}$, thus contradicting the fact $\bar{y}$ is a solution of \eqref{Eq:ContinuousEE2}.
\end{IEEEproof}
Despite having a convex feasibility set, Problem \eqref{Eq:ContinuousEE2} is still challenging to solve, since the numerator and denominator of \eqref{Eq:ContinuousEE2a} are not concave and convex functions, respectively, which prevents one from using fractional programming techniques. However, recalling Lemma \ref{Prop:G1}, fractional programming can be used if $y$ is fixed. More precisely, for any fixed $y$, Problem \eqref{Eq:ContinuousEE2} is an instance of a so-called pseudo-concave maximization problem, in which the fraction to maximize has a concave numerator and an affine denominator, and thus can be solved with limited complexity by any fractional programming method, such as the popular Dinkelbach's method \cite{NonlinearFracProg}. Moreover, from \eqref{Eq:ContinuousEE2e}, it must hold that 
\beq\label{Eq:yInterval}
y\in\left[\frac{d}{\beta},B_{max}\log\left(1+\frac{P_{max}|h_{F}|^{2}}{B_{max}N_{0}}\right)\right]\;.
\eeq
Based on these considerations, Problem \eqref{Eq:ContinuousEE2e} can be solved by performing a line search over $y$ in the interval given by \eqref{Eq:yInterval}, and solving, for each considered value $\tilde{y}$, the corresponding pseudo-concave maximization problem as follows
\begin{subequations}\label{Eq:yEE}
\begin{align}
&\ds\max_{p,B,p_{F}} \frac{(\beta-\frac{d}{\tilde{y}})B\log\left(1+\frac{pc}{B}\right)}{\beta\mu p+P_{c}+\frac{d}{\tilde{y}}(\mu_{F}p_{F}-\mu p)}
\label{Eq:yEEa}\\
&\;\text{s.t.}\;p+p_{F}\leq P_{max}\;,0\leq B\leq B_{max}\label{Eq:yEEb}\\
&\quad\;\;p\geq 0\;,\;p_{F}\geq0\label{Eq:EEd}\\
&\quad\;\; (B_{max}-B)\log\left(1+\frac{p_{F}|h_{F}|^{2}}{(B_{max}-B)N_{0}}\right)\geq \tilde{y} \label{Eq:yEEe}
\end{align}
\end{subequations}
Thus we have Algorithm \ref{Alg:OptEE}, wherein $\text{EE}_{m}$ denotes the value of \eqref{Eq:yEEa} obtained at the $m$-th iteration.

\begin{algorithm}\caption{EE Maximization}
\begin{algorithmic}\small\label{Alg:OptEE}
\STATE \texttt{Set} $M>0$ \texttt{and compute}
\beq
\Delta=\frac{B_{max}\log\left(1+\frac{P_{max}|h_{F}|^{2}}{B_{max}N_{0}}\right)-\frac{d}{\beta}}{M}
\eeq
\FOR{$m=1,\ldots,M$}
\STATE $\tilde{y}_{m}=\frac{d}{\beta}+(m-1)\Delta$;
\STATE \texttt{Solve} \eqref{Eq:yEE} \hspace{-0.2cm}\texttt{ and compute }\hspace{-0.2cm} $\text{EE}_{m}(p_{m}^{*},p_{m,F}^{*},B_{m}^{*},\tilde{y}_{m})$
\ENDFOR
\STATE \texttt{Compute} $m^{*}=\text{argmax}\;\text{EE}_{m}$;
\STATE \texttt{Output} $p_{m^{*}}^{*},p_{m^{*},F}^{*},B_{m^{*}}^{*},B_{m^{*},F}^{*}=B_{max}-B_{m^{*}}^{*}$;
\end{algorithmic}
\end{algorithm}

\subsection{Rate-EE optimization}\label{Sec:BiObjectiveOpt}
This section focuses on characterizing the rate-energy Pareto-optimal frontier of the bi-objective problem that has as objectives the system rate and the energy efficiency.

To begin with, since $\bPhi,\bq,\bw$ affect only the numerator of the energy efficiency, which coincides with the rate, we can plug any of the allocations of $\bPhi,\bq,\bw$ developed in Section \ref{Sec:PhaseOpt} into the rate and the energy efficiency functions, which yields 
\begin{subequations}\label{Eq:Trade-offJoint}
\begin{align}
&\max_{p,p_{F},B}\big\{R(p,p_{F},B,\bPhi^{\text{opt}},\bq^{\text{opt}},\bw^{\text{opt}}),\label{Eq:aTrade-offJoint}\\
&\hspace{3cm}\text{EE}(p,p_{F},B,\bPhi^{\text{opt}},\bq^{\text{opt}},\bw^{\text{opt}})\big\}\notag\\
&\;\text{s.t.}\;p+p_{F}\!\leq\! P_{max}\;,0\!\leq \!B\!\leq \!B_{max}, p\geq 0\;,p_{F}\geq 0\label{Eq:bTrade-offJoint}\\
&\quad\;\;\frac{d}{(B_{max}-B)\log\left(1+\frac{p_{F}|h_{F}|^{2}}{(B_{max}-B)N_{0}}\right)}\leq \beta\label{Eq:fTrade-offJoint}\;,
\end{align}
\end{subequations}
where we have already exploited the fact that at the optimum it must hold $B+B_{F}=B_{max}$. With respect to the other variables, on the other hand, the rate and energy efficiency are in general maximized by different resource allocations. Clearly, this is the scenario in which Problem \eqref{Eq:Trade-offJoint} is of interest, because otherwise no trade-off would exist between the two functions, and the solution of Problem \eqref{Eq:Trade-offJoint} would be trivially equal to the common maximizer of the rate and of the energy efficiency. 

The most widely-used solution concept for bi-objective problems like \eqref{Eq:Trade-offJoint} is that of Pareto-optimality. A Pareto-optimal solution of \eqref{Eq:aTrade-offJoint} is a point lying on the so-called Pareto-frontier of the problem, defined as the set of resource allocations for which it is not possible to further increase either one of the two objectives, without decreasing the other objective. To elaborate further, let us denote by $R_{opt}$ and $\text{EE}_{opt}$ the maximum rate and energy efficiency that can be computed as shown in Sections \ref{Sec:RateOpt} and \ref{Sec:EEOptimization}, respectively. Then, we also denote by $R_{\text{EE}_{opt}}$ the rate obtained with the resource allocation that maximizes the energy efficiency, and by $\text{EE}_{R_{opt}}$ the energy efficiency obtained with the resource allocation that maximizes the rate. Then, it follows that the extreme points of the Pareto-frontier in the $R-\text{EE}$ plane are $(R_{opt},\text{EE}_{R_{opt}})$ and $(R_{\text{EE}_{opt}},\text{EE}_{opt})$. As expected, this also shows that the Pareto-frontier degenerates into a single point when the rate and the energy efficiency admit the same maximizer. Instead, in general a non-trivial Pareto-frontier exists for \eqref{Eq:Trade-offJoint}, which provides all optimal trade-off points between the rate and the energy efficiency. Focusing on this scenario, multi-objective optimization theory provides several approaches to compute all Pareto-optimal points of a multi-objective problem. One of the most widely-used methods is the maximization of the minimum between a weighted combination of the objectives. As for Problem \eqref{Eq:Trade-offJoint}, introducing the auxiliary variable $y$ defined in \eqref{Eq:yAux}, the max-min approach leads to the problem:
\begin{subequations}\label{Eq:Trade-offJointMaxmin}
\begin{align}
&\max_{p,p_{F},B,y}\!\min\!\Bigg\{\!\alpha \left(R(p,y,B,\!\bPhi^{\text{opt}},\!\bq^{\text{opt}},\!\bw^{\text{opt}})\!-\!R_{\text{opt}}\right),\label{Eq:aTrade-offJointMaxmin}\\
&\hspace{1.7cm}(1\!-\!\alpha)\left(\frac{R(p,y,B,\!\bPhi^{\text{opt}},\!\bq^{\text{opt}},\!\bw^{\text{opt}})}{\beta\mu p+P_{c}+\frac{d}{y}(\mu_{F}p_{F}-\mu p)}\!-\!\text{EE}_{\text{opt}}\!\right)\!\Bigg\}\notag\\
&\;\text{s.t.}\;p+p_{F}\!\leq \!P_{max}\;,0\!\leq \!B\!\leq\! B_{max}\!\;,p\!\geq 0\;,p_{F}\geq 0\label{Eq:bTrade-offJointMaxmin}\\
&\quad\;\;\frac{d}{\beta}\!\leq\! y\!\leq\! (B_{max}-B)\log\left(1\!+\!\frac{p_{F}|h_{F}|^{2}}{(B_{max}\!-\!B)N_{0}}\right)\label{Eq:fTrade-offJointMaxmin}
\end{align}
\end{subequations}
wherein we have plugged in the expression of the energy efficiency, with $R(p,y,B,\bPhi^{\text{opt}},\bq^{\text{opt}},\bw^{\text{opt}})=(\beta-\frac{d}{y})B\log\left(1+\frac{pc}{B}\right)$, $\alpha$ is a non-negative parameters that weighs the relative importance between the rate and the energy efficiency, while $\text{R}_{\text{opt}}$ and $\text{EE}_{\text{opt}}$ are the maximum of the rate and of the energy efficiency, respectively. For any $\alpha\in(0,1)$, \eqref{Eq:Trade-offJointMaxmin} has at least one solution that is Pareto-optimal for \eqref{Eq:Trade-offJointMaxmin} \cite[Theorem 3.4.3]{MOBook}, and solving \eqref{Eq:Trade-offJointMaxmin} for all $\alpha\in(0,1)$ yields all the points on the Pareto-frontier of \eqref{Eq:Trade-offJoint} \cite[Theorem 3.4.5]{MOBook}. Also, the two extreme points $\alpha=1$ and $\alpha=0$ correspond to the single-objective maximization of the rate and of the energy efficiency. In order to solve \eqref{Eq:Trade-offJointMaxmin}, we consider its equivalent reformulation in epigraph form, namely
\begin{subequations}\label{Eq:Trade-offJointMaxminEpi}
\begin{align}
&\max_{p,p_{F},B,y,t}\; t\label{Eq:aTrade-offJointMaxminEpi}\\
&\;\text{s.t.}\;p+p_{F}\leq P_{max}\label{Eq:bTrade-offJointMaxminEpi}\\
&\quad\;\;0\leq B\leq B_{max}\;,p\geq 0\;,\;p_{F}\geq0\label{Eq:cTrade-offJointMaxminEpi}\\
&\quad\;\;\frac{d}{\beta}\!\leq\! y\!\leq\! (B_{max}-B)\log\left(1+\frac{p_{F}|h_{F}|^{2}}{(B_{max}-B)N_{0}}\right)\label{Eq:fTrade-offJointMaxminEpi}\\
&\quad\;\; \left(\beta-\frac{d}{y}\right)B\log\left(1+\frac{pc}{B}\right)\geq \frac{t}{\alpha}+ R_{\text{opt}}\\
&\quad\;\; \left(\beta-\frac{d}{y}\right)B\log\left(1+\frac{pc}{B}\right)\geq \left(\frac{t}{1\!-\!\alpha}\!+\!\text{EE}_{\text{opt}}\right)\times \notag\\
&\hspace{3cm}\left(\beta\mu p+P_{c}\!+\!\frac{d}{y}(\mu_{F}p_{F}\!-\!\mu p)\right)
\end{align}
\end{subequations}
Solving \eqref{Eq:Trade-offJointMaxmin} is challenging due to the presence of the variable $y$. However, for any fixed $y$, \eqref{Eq:Trade-offJointMaxmin} can be conveniently solved by employing  the bisection algorithm over $t$, since all constraint functions are convex in all other variables. Specifically, observing that $y$ must lie in the interval defined by \eqref{Eq:yInterval}, Problem \eqref{Eq:Trade-offJointMaxminEpi} can be solved by performing a line search over $y$, solving in each iteration the following problem with $y=\tilde{y}$ lying in in the interval defined by \eqref{Eq:yInterval}: 
\begin{subequations}\label{Eq:Trade-offJointMaxminEpi2}
\begin{align}
&\max_{p,p_{F},B,t}\; t\label{Eq:aTrade-offJointMaxminEpi2}\\
&\;\text{s.t.}\;p+p_{F}\leq P_{max}\label{Eq:bTrade-offJointMaxminEpi2}\\
&\quad\;\;0\leq B\leq B_{max}\;,p\geq 0\;,\;p_{F}\geq0\label{Eq:cTrade-offJointMaxminEpi2}\\
&\quad\;\; (B_{max}-B)\log\left(1+\frac{p_{F}|h_{F}|^{2}}{(B_{max}-B)N_{0}}\right)\geq \tilde{y}\label{Eq:fTrade-offJointMaxminEpi2}\\
&\quad\;\; \left(\beta-\frac{d}{\tilde{y}}\right)B\log\left(1+\frac{pc}{B}\right)\geq \frac{t}{\alpha}+ R_{\text{opt}}\\
&\quad\;\; \left(\beta-\frac{d}{\tilde{y}}\right)B\log\left(1+\frac{pc}{B}\right)\geq \left(\frac{t}{1\!-\!\alpha}\!+\!\text{EE}_{\text{opt}}\right)\times\notag\\
&\hspace{3cm}\left(\beta\mu p+P_{c}\!+\!\frac{d}{\tilde{y}}(\mu_{F}p_{F}\!-\!\mu p)\right)
\end{align}
\end{subequations}
Problem \eqref{Eq:Trade-offJointMaxmin} can be solved similarly as in Algorithm \ref{Alg:OptEE}. Formally, this yields Algorithm \ref{Alg:TradeOff}.
\begin{algorithm}\caption{Rate-EE Maximization}
\begin{algorithmic}\small\label{Alg:TradeOff}
\STATE \texttt{Set} $M>0$ \texttt{and compute} $\Delta=\frac{B_{max}\log\left(1+\frac{P_{max}|h_{F}|^{2}}{B_{max}N_{0}}\right)-\frac{d}{\beta}}{M}$
\FOR{$m=1,\ldots,M$}
\STATE $\tilde{y}_{m}=\frac{d}{\beta}+(m-1)\Delta$;
\STATE \texttt{Solve} \eqref{Eq:Trade-offJointMaxminEpi2} \texttt{by bisection over} $t$ \hspace{-0.2cm}\texttt{ and compute}
\begin{align}
F_{m}\!=\!&\min\!\Bigg\{\!\alpha\!\left(R(p^{*}\!,p_{F}^{*}\!,B^{*}\!,\bPhi^{\text{opt}},\!\bq^{\text{opt}},\!\bw^{\text{opt}},)\!-\!R_{\text{opt}}\right),\\
&\hspace{1.5cm}(1\!-\!\alpha)\!\left(\text{EE}(p^{*}\!,p_{F}^{*}\!,B^{*},\!\bPhi^{\text{opt}},\!\bq^{\text{opt}},\!\bw^{\text{opt}})\!-\!\text{EE}_{\text{opt}}\right)\!\!\!\Bigg\}\notag
\end{align}
\ENDFOR
\STATE \texttt{Compute} $m^{*}=\text{argmax}\;F_{m}$;
\STATE \texttt{Output} $p_{m^{*}}^{*},p_{m^{*},F}^{*},B_{m^{*}}^{*},B_{m^{*},F}^{*}=B_{max}-B_{m^{*}}^{*}$;
\end{algorithmic}
\end{algorithm}

\section{Optimality properties and computational complexity}
This section analyzes the properties and complexity of the proposed optimization algorithms. The algorithms developed in Sections \ref{Sec:UpperPhase} and \ref{Sec:LowerPhase} are discussed in Section \ref{Sec:PhaseOptComplexity}, while those developed in Section \ref{Sec:ContinuousRate} are discussed in Section \ref{Sec:PowerOptComplexity}.
\subsection{Algorithms for the optimization of $\bPhi$, $\bq$, $\bw$}\label{Sec:PhaseOptComplexity}
The algorithms for the optimization of the \gls{ris} phase shifts, the transmit beamforming, and the receive vector introduced in Sections \ref{Sec:UpperPhase} and \ref{Sec:LowerPhase} are based on the use of upper and lower bounds of the receive \gls{sinr}. As a result, in general they are not globally optimal. Nevertheless, they achieve global optimality whenever the rank of $\bH$ and $\bG$ are equal to one. Indeed, in this case both the upper-bound in Section \ref{Sec:UpperPhase} and the lower-bound in Section \ref{Sec:LowerPhase} are tight, because when $r_{G}=r_{H}=1$, the vectors $\bq$ and $\bw$ reduce to scalars. The case of rank-one channles includes two notable special cases:
\begin{itemize}
\item The case in which a single-antenna is used at the transmit and receive side.
\item The use of mmWave communications, which, in many cases, leads to rank-one channels as all energy is focused in a pencil-beam transmission.
\end{itemize} 
In general, as we have explained at the beginning of Section \ref{Sec:PhaseOpt}, jointly optimizing $\bPhi$, $\bq$, and $\bw$ in a globally optimal way is computationally prohibitive due to the lack of a tractable and closed-form expression for the dominant singular value of the matrix $\bA=\bG\bPhi\bH$. As a result, the global joint optimization of $\bPhi$, $\bq$, and $\bw$ would require an exhaustive search in an $NN_{T}N_{R}$-dimensional space. This justifies the use of possibly sub-optimal optimization methods, among which the state-of-the-art approach is the alternating optimization algorithm reviewed in Section \ref{Sec:AOPhase}. Here, we show that the two novel approaches developed in Sections \ref{Sec:UpperPhase} and \ref{Sec:LowerPhase} require a lower computational complexity than alternating optimization. 

To elaborate, alternating optimization is an iterative approach, which requires to compute, in each iteration of the algorithm, the \gls{svd} of the matrix $\bG\bPhi\bH$, as well as the vectors $\bg_{w}=\bG^{H}\bw$ and $\bh_{q}=\bH\bq$ to set the \gls{ris} phase shifts to $\phi_{n}=-\angle{\{\bg_{w}^{*}(n)\bh_{q}(n)\}}$, for all $n=1,\ldots,N$. Thus, if $N_{it}$ is the number of iterations until the alternating optimizations converges, the above operations are to be executed $N_{it}$ times. Instead, the advantage of our proposed methods is that they are not iterative, but are based on closed-form optimization results. Specifically, both methods from Sections \ref{Sec:UpperPhase} and \ref{Sec:LowerPhase} require the computation of a single \gls{svd} and a single \gls{ris} phase adjustment of the form $\phi_{n}=-\angle{\{\bg_{w}^{*}(n)\bh_{q}(n)\}}$. In addition, the method developed in Section \ref{Sec:UpperPhase} requires two $\text{argmax}(\cdot)$ searches over finite sets of size $r_{G}$ and $r_{H}$, respectively,  while the method developed in Section \ref{Sec:LowerPhase} requires computing the matrix $\sum_{n=1}^{N}\bg_{n}\bh_{n}^{T}$. Again, all of these additional operations are to be executed only once, and their complexity is negligible compared to that of performing an \gls{svd}. In summary, since the proposed algorithms are not iterative, but are based on closed-form optimization expressions, they reduce the complexity compared to alternating optimization by a factor $N_{it}$. Moreover, Section \ref{Sec:Numerics} will numerically show that the proposed  methods perform very close to alternating optimization. 

\subsection{Algorithms for the optimization of $p$, $p_F$, $B$, $B_F$}\label{Sec:PowerOptComplexity}
All algorithms developed for the optimization of the transmit and feedback power and bandwidths are globally optimal and require a limited computational complexity. Specifically:
\begin{itemize}
\item Rate optimization has been recast as a concave maximization, which is optimally solvable with polynomial complexity in the number of optimization variables \cite{boyd2004convex}.
\item The energy efficiency maximization problem has been reformulated as a pseudo-concave maximization problem upon fixing the value of the auxiliary variable $y$. Thus, energy efficiency maximization can be optimally performed by a scalar line search over $y$ and by solving a pseudo-concave maximization problem for each considered value of $y$. Recalling that polynomial complexity algorithms exist to solve pseudo-concave maximizations \cite{ZapNow15}, the complexity of energy efficiency maximization is polynomial in the number of optimization variables, and linear in the number of points $M$ used for the line search. 
\item The bi-objective problem of rate and energy efficiency maximization has been reformulated as the feasibility test in \eqref{Eq:Trade-offJointMaxminEpi}, that can be optimally solved by a sequence of feasibility tests of the form of Problem \eqref{Eq:Trade-offJointMaxminEpi2}, which become convex when fixing the variable $y$. Thus, the complexity of rate and energy efficiency bi-objective maximization is polynomial in the number of optimization variables, and linear in the number of points $M$ used for the line search. Moreover, the optimal parameter $t$ is determined by a bisection search, which requires solving \eqref{Eq:Trade-offJointMaxminEpi2} $\log_{2}\left\lceil\frac{U-L}{\epsilon}\right\rceil$ times, with $U$ and $L$ the initialization of the bisection method, and $\epsilon$ the accuracy of the bisection search \cite{boyd2004convex}.
\end{itemize}

\section{Numerical Results}\label{Sec:Numerics}
Consider the system model described in Section \ref{Sec:SysModel}, with system parameters set as in Table \ref{Tab:NetParameters}. 
\begin{table}\small
\centering
  \begin{tabular}{ | c | c | c | c | c | c | c | c | c | c |}
  \hline
    $P_{max}/ P_{c,0} / P_{c,n}$ & $B_{max}$ & $N_0$ & $\mu / \mu_F$ & $b_F$\\
    \hline
    45 / 45 / 10\! dBm  & 100 \!MHz  & -174\! dBm/Hz  & 1 / 1 & 16 \!\!\! bit \\
    \hline
  \end{tabular}\caption{Network parameters}\label{Tab:NetParameters}
\end{table}
For all $nt=1,\ldots,N_{T}$, $nr=1,\ldots,N_{R}$, $n=1,\ldots,N$, each product channel is generated as $h_{nt,n}g_{n,nr}=\frac{\alpha_{h}\alpha_{g}}{\sqrt{\beta}}$, wherein $\alpha_{h}$ and $\alpha_{g}$ are realizations of two independent complex circularly symmetric standard Gaussian variable, while $\beta$ accounts for the overall path-loss and shadowing effects from the transmitter to the \gls{ris} and from the \gls{ris} to the receiver\footnote{Rayleigh fading is a suitable case study in scenarios in which the location of the \gls{ris} can not be optimized and the existence of a strong line-of-sight component can not be guaranteed. This is the case when the \glspl{ris} are randomly deployed, e.g., on spatial blockages whose locations are not under the control of the system designer.}. In our simulations, we set $\beta=10^{0.1*\beta_{dB}}$, with $\beta_{dB}=110$. A similar model is used for the feedback channel $h_{F}$.
 
\begin{figure}[h]
 \centering
 \subfigure[Spectral efficiency for $T_{0}=0.8\,\mu s$]
   {    \psfrag{N}[c][c][0.9]{$N$}
    \psfrag{Spectral efficiency [bits/s/Hz]}[c][c][0.9]{$SE$ [bits/s/Hz]}
    \includegraphics[width=0.4\textwidth,height=0.2\textheight]{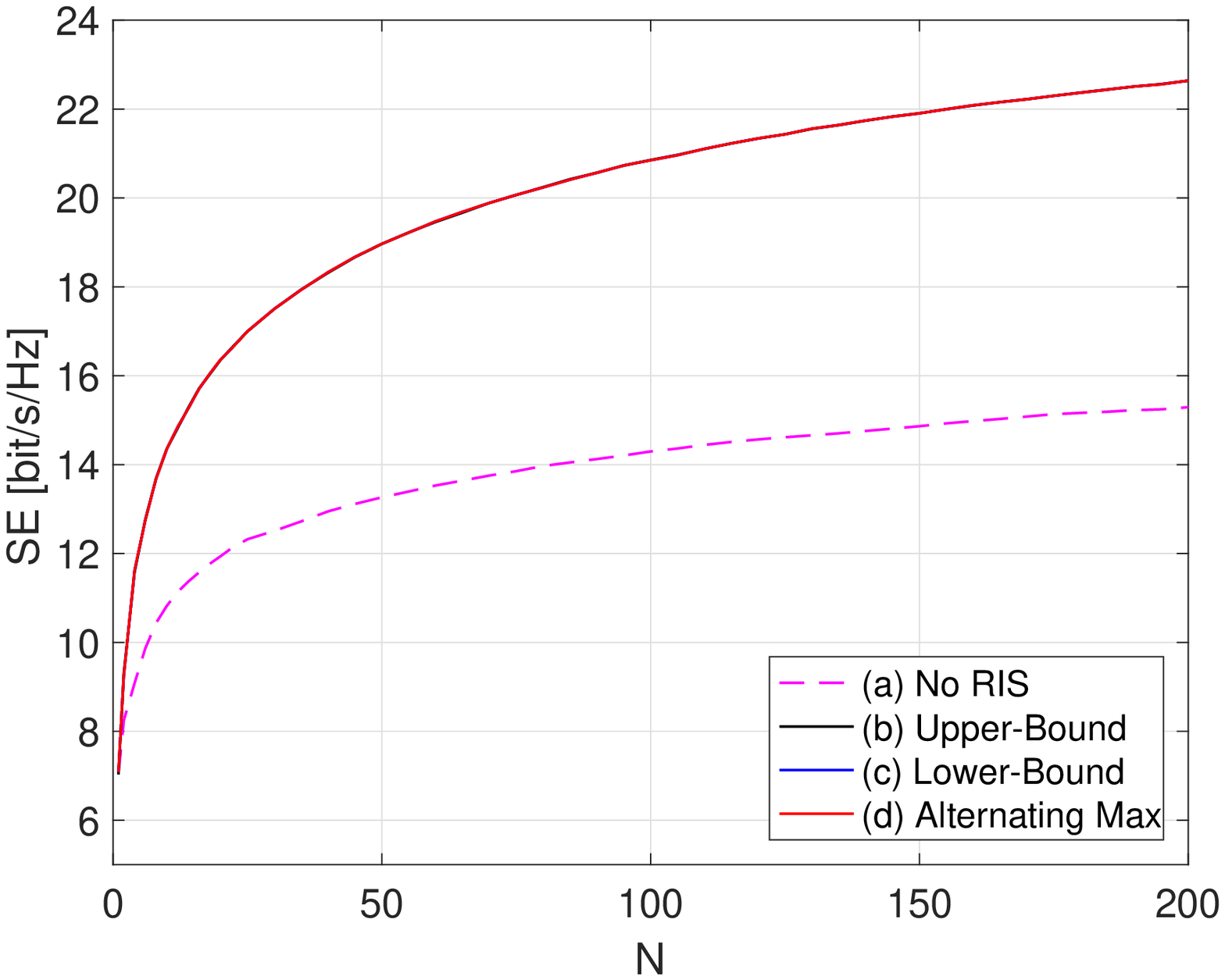}\label{fig:maxR_NT1NR1_LargeT0}}
 \subfigure[Spectral efficiency for $T_{0}=0.15\,\mu s$]
   {    \psfrag{N}[c][c][0.9]{$N$}
    \psfrag{Spectral efficiency [bits/s/Hz]}[c][c][0.9]{$SE$ [bits/s/Hz]}
    \includegraphics[width=0.4\textwidth,,height=0.2\textheight]{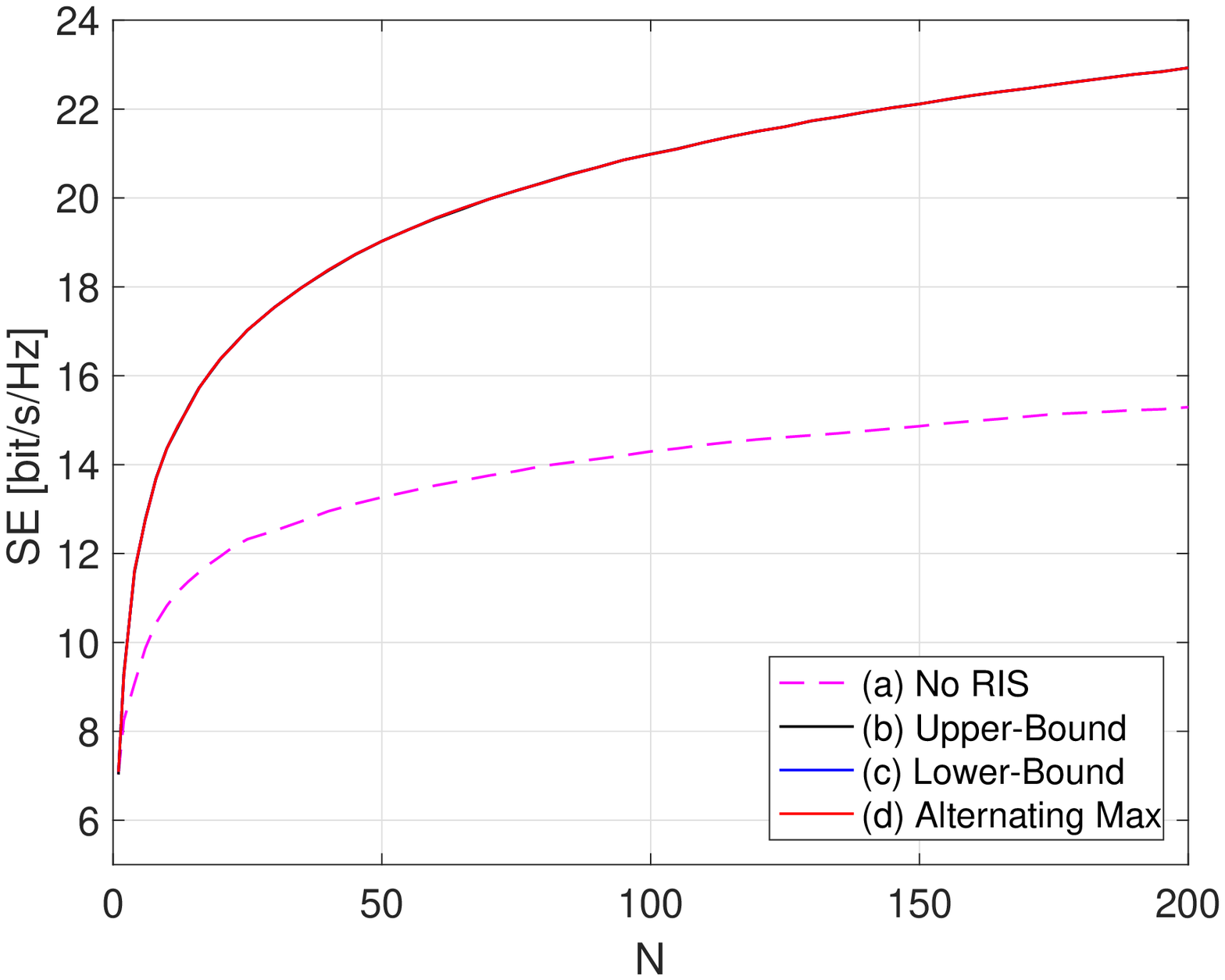} \label{fig:maxR_NT1NR1_SmallT0}}
 \caption{Spectral efficiency as a function of $N$ for $N_T=N_R=1$.}\label{fig:maxR_NT1NR1}\vspace{-0.35cm}
 \end{figure}

Figs. 2-9 assume that the overhead model from Section \ref{Sec:Overhead}, Case (a), is employed.
Figure \ref{fig:maxR_NT1NR1} plots the maximum rate in \eqref{Eq:BitsTx} (normalized by $B_{max}$) versus $N$, with $N_T=N_R=1$, $T_{0}=0.8\,\mu s$ (Fig. \ref{fig:maxR_NT1NR1}-a), and $T_{0}=0.15\,\mu s$ (Fig. \ref{fig:maxR_NT1NR1}-b), for:
\begin{itemize}
\item [(a)] $p,p_{F},B,B_{F}$ obtained from the optimal method from Section \ref{Sec:ContinuousRate}, with $\bPhi=\bI_{N}$, and $\bq$, $\bw$ chosen as the dominant right and left eigenvectors of $\bA=\bH\bPhi\bG$. Thus, the RIS simply reflects the signal without any phase manipulation. It is worth noting that in this case there is no need to configure the phase shifts of the \gls{ris}, and, therefore, the total overhead is much reduced. In particular, the numerical results that correspond to this case are obtained by setting $T_F=0$ and $T_E = N_T N_R T_0$. 
\item [(b)] $p,p_{F},B,B_{F}$ obtained from the optimal method from Section \ref{Sec:ContinuousRate} and  $\bPhi_{up},\bq_{up},\bw_{up}$ obtained from the maximization of the upper-bound derived in Section  \ref{Sec:UpperPhase}.
\item [(c)] $p,p_{F},B,B_{F}$ obtained from the optimal method from Section \ref{Sec:ContinuousRate} and  $\bPhi_{low},\bq_{low},\bw_{low}$ obtained from the maximization of the lower-bound derived in Section  \ref{Sec:LowerPhase}.
\item [(d)] $p,p_{F},B,B_{F}$ obtained from the optimal method from Section  \ref{Sec:ContinuousRate}, and  $\bPhi_{alt},\bq_{alt},\bw_{alt}$ obtained from the alternating maximization Algorithm \ref{Alg:OptPhase} in  Section \ref{Sec:AOPhase}.
\end{itemize} 
The results in Figure \ref{fig:maxR_NT1NR1} indicate that the proposed schemes are able to outperform the case in which no RIS optimization is performed, which shows that the use of \glspl{ris} can significantly improve the system performance, even if the overhead for channel estimation and system configuration is taken into account. Moreover, it is observed that the proposed closed-form Schemes (b) and (c) offer similar performance as  alternating optimization, which instead requires the implementation of an iterative numerical algorithm. Indeed, we recall that when $N_{T}=N_{R}=1$, Schemes (b) and (c) are provably optimal. 

\begin{figure}[h]
 \centering
 \subfigure[Spectral efficiency for $T_{0}=0.8\,\mu s$]
   {    \psfrag{N}[c][c][0.9]{$N$}
    \psfrag{Spectral efficiency [bits/s/Hz]}[c][c][0.9]{$SE$ [bits/s/Hz]}
    \includegraphics[width=0.4\textwidth,height=0.2\textheight]{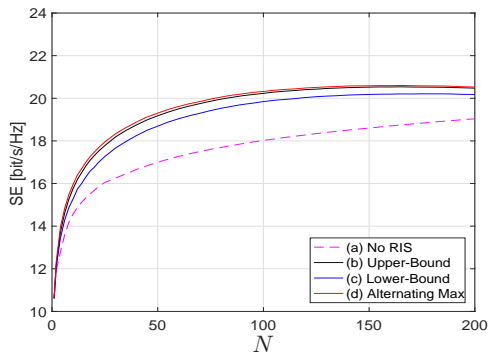}\label{fig:maxR_NT1NR8_LargeT0}}
 \subfigure[Spectral efficiency for $T_{0}=0.15\,\mu s$]
   {    \psfrag{N}[c][c][0.9]{$N$}
    \psfrag{Spectral efficiency [bits/s/Hz]}[c][c][0.9]{$SE$ [bits/s/Hz]}
    \includegraphics[width=0.4\textwidth,height=0.2\textheight]{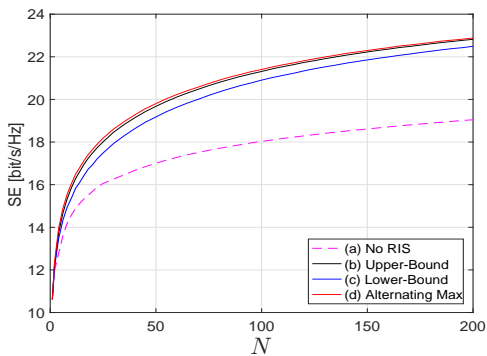} \label{fig:maxR_NT1NR8_SmallT0}}
 \caption{Spectral efficiency as a function of $N$ for $N_T=1$, $N_R=8$.}\label{fig:maxR_NT1NR8}\vspace{-0.35cm}
 \end{figure}

In order to show the impact of the overhead that is necessary to operate \gls{ris}-empowered wireless networks, Figure \ref{fig:maxR_NT1NR8} considers a similar scenario as in Figure \ref{fig:maxR_NT1NR1}, with the only difference that the number of receive antennas is set to $N_{R}=8$, which significantly increases the amount of feedback data. As a result, it is observed that the gap between Schemes (b), (c), (d), which optimize the phase shifts of the \gls{ris}, and Scheme (a) without \gls{ris} optimization, gets smaller, since not optimizing the phases allows one to dispense with the overhead to obtain the channels $\bH$ and $\bG$ for each individual phase shift. Also, the gap is smaller when a larger $T_{0}$ is considered, since a longer time is needed for channel estimation and feedback. Moreover, it is interesting to observe that Scheme (b) performs similar to alternating optimization, despite requiring a much lower computational complexity thanks to the fact that it provides a closed-form allocation. On the other hand, Scheme (d) shows a slight gap compared to Schemes (b) and (d). 

\begin{figure}[h]
 \centering
 \subfigure[Spectral efficiency for large $T_{0}=0.8\,\mu s$]
   {    \psfrag{N}[c][c][0.9]{$N$}
    \psfrag{Spectral efficiency [bits/s/Hz]}[c][c][0.9]{$SE$ [bits/s/Hz]}
    \includegraphics[width=0.4\textwidth,height=0.2\textheight]{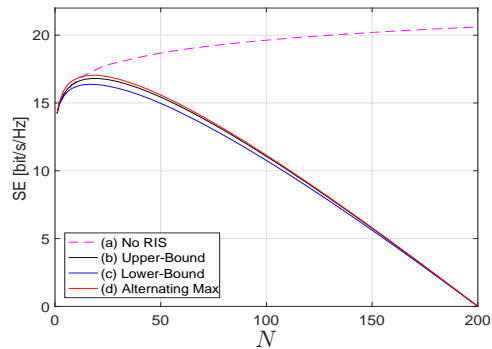}\label{fig:maxR_NT8NR8_LargeT0}}
 \subfigure[Spectral efficiency for small $T_{0}=0.15\,\mu s$]
   {    \psfrag{N}[c][c][0.9]{$N$}
    \psfrag{Spectral efficiency [bits/s/Hz]}[c][c][0.9]{$SE$ [bits/s/Hz]}
    \includegraphics[width=0.4\textwidth,height=0.2\textheight]{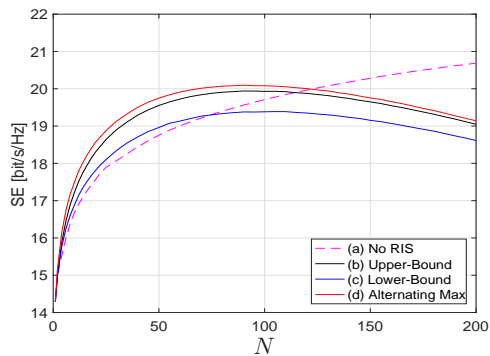} \label{fig:maxR_NT8NR8_SmallT0}}
 \caption{Spectral efficiency as a function of $N$ for $N_T=8$, $N_R=8$.}\label{fig:maxR_NT8NR8}\vspace{-0.35cm}
 \end{figure}

The trend displayed in Figure \ref{fig:maxR_NT1NR8} becomes even more significant in Figure \ref{fig:maxR_NT8NR8}, where the number of antennas is further increased by considering $N_{T}=N_{R}=8$. In this case, Scheme (a) which does not require any overhead for the optimization of the \gls{ris} phase shifts, outperforms the system setup in the presence of an \gls{ris}, when $T_{0}=0.8\,\mu s$, i.e., when a longer time is used for channel estimation. Instead, when a shorter channel estimation time is used, i.e., when $T_{0}=0.15\,\mu s$, performing radio resource allocation is still beneficial up to $N=130$, whereas not using an \gls{ris}  becomes better for higher values of $N$. Moreover, also in this case Schemes (b) and (d) perform very similarly, while Scheme (c) exhibits a slight gap. 

The obtained results motivate the use of RISs in scenarios with a low number of transmit and receive antennas, especially for large $N$. Indeed, for any additional antenna that is deployed, $N$ new channels must be estimated and the optimized phases need to be communicated to the \gls{ris}. Comparing the performance of the optimized schemes in Figures \ref{fig:maxR_NT1NR1} and \ref{fig:maxR_NT1NR8} reveals that deploying a moderate number of antennas does not lead to improved performance. Indeed, the presence of an \gls{ris} may make transmit beaforming and receive combining not necessary. This finding agrees with recent results from \cite{Arun2019}.

\begin{figure}[!h]
 \centering
 \subfigure[EE for $T_{0}=0.8\,\mu s$, $P_{0}=2.5\,\textrm{mW}$]
   {\psfrag{N}[c][c][0.9]{$N$}
    \psfrag{EE [Mbits/Joule]}[c][c][0.9]{$EE $ [Mbits/Joule]}
    \includegraphics[width=0.4\textwidth,height=0.2\textheight]{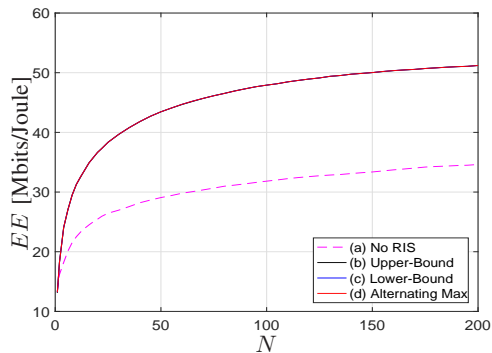}\label{fig:maxEE_NT1NR1_LargeT0}}
\subfigure[EE for $T_{0}=0.15\,\mu s$, $P_{0}=0.5\,\textrm{mW}$]
   {\psfrag{N}[c][c][0.9]{$N$}
    \psfrag{EE [Mbits/Joule]}[c][c][0.9]{$EE $ [Mbits/Joule]}
    \includegraphics[width=0.4\textwidth,height=0.2\textheight]{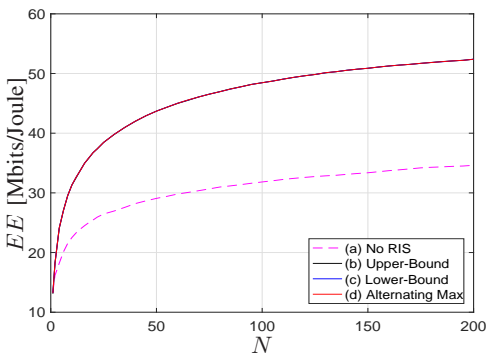} \label{fig:maxEE_NT1NR1_SmallT0}}
 \caption{Achieved EE in [Mbit/Joule] as a function of $N$ for $N_T=N_R=1$.}\label{fig:maxEE_NT1NR1}\vspace{-0.35cm}
 \end{figure}

\begin{figure}[!h]
 \centering
 \subfigure[EE for $T_{0}=0.8\,\mu s$, $P_{0}=2.5\,\textrm{mW}$]
   {    \psfrag{N}[c][c][0.9]{$N$}
    \psfrag{EE [Mbits/Joule]}[c][c][0.9]{$EE $ [Mbits/Joule]}
    \includegraphics[width=0.4\textwidth,height=0.2\textheight]{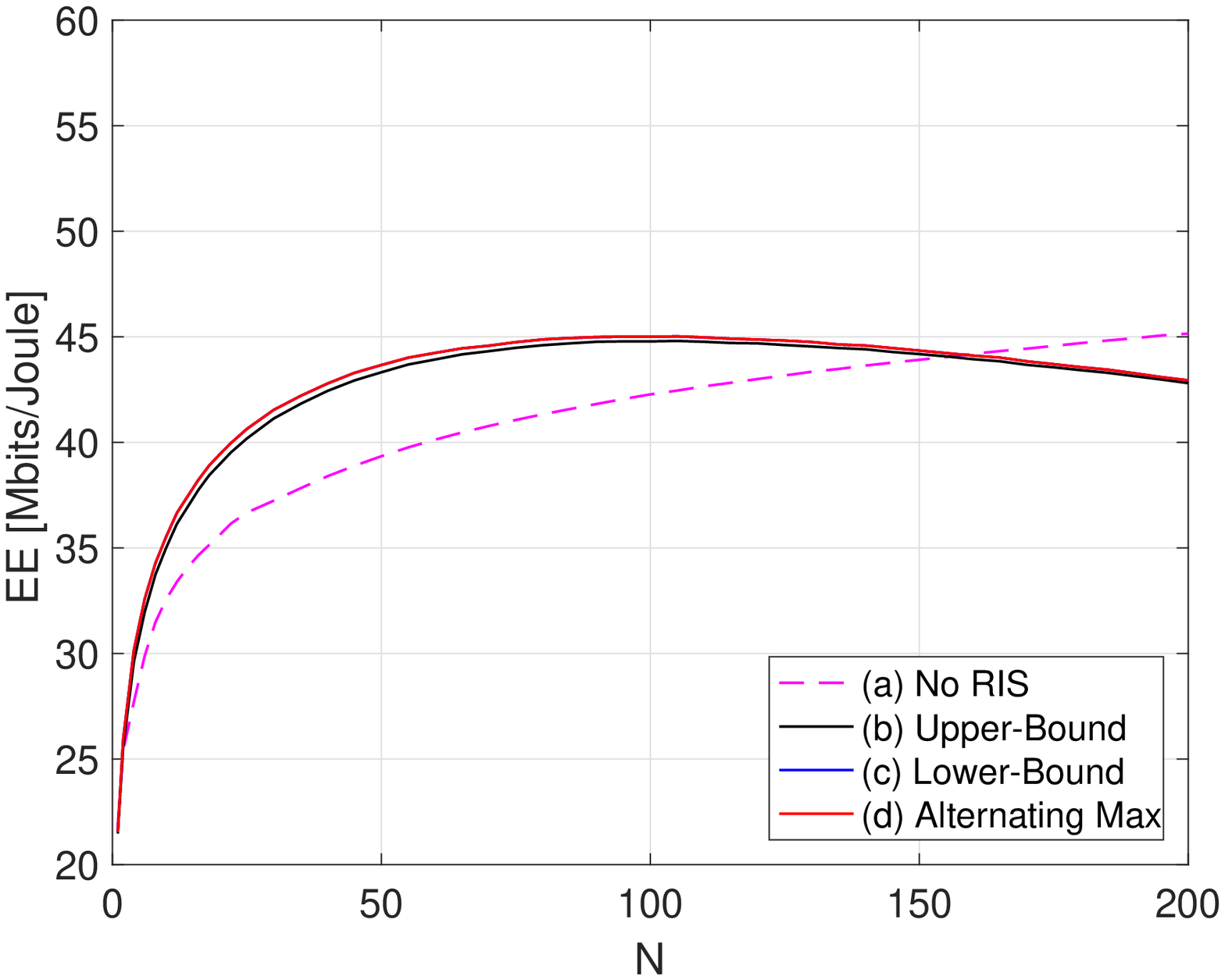}\label{fig:maxEE_NT1NR8_LargeT0}}
 \subfigure[EE for $T_{0}=0.15\,\mu s$, $P_{0}=0.5\,\textrm{mW}$]
   {    \psfrag{N}[c][c][0.9]{$N$}
    \psfrag{EE [Mbits/Joule]}[c][c][0.9]{$EE $ [Mbits/Joule]}
    \includegraphics[width=0.4\textwidth,height=0.2\textheight]{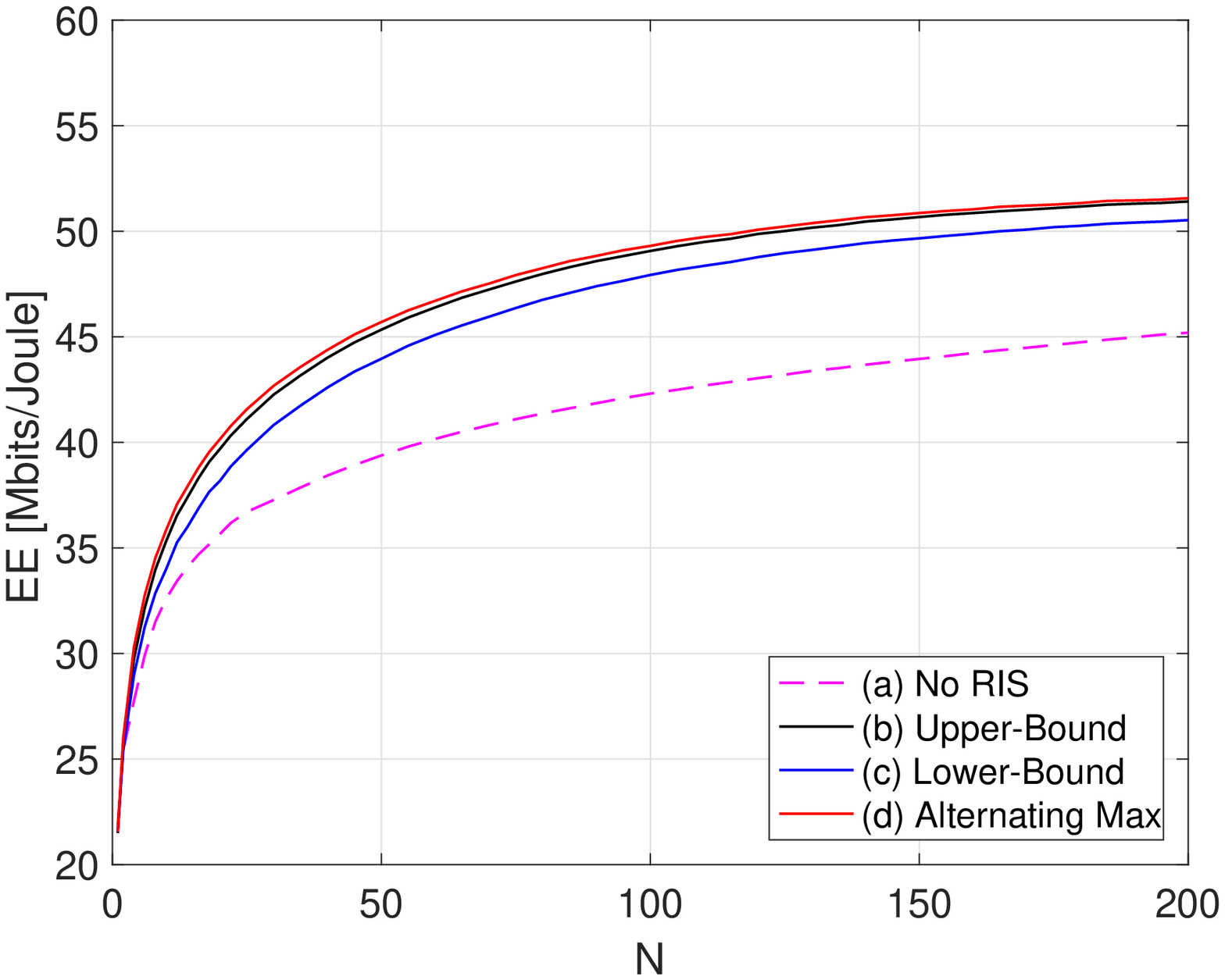} \label{fig:maxEE_NT1NR8_SmallT0}}
 \caption{Achieved EE in [Mbit/Joule] as a function of $N$ for $N_T=1$, $N_R=8$.}\label{fig:maxEE_NT1NR8}\vspace{-0.35cm}
 \end{figure}

\begin{figure}[!h]
 \centering
 \subfigure[EE for $T_{0}=0.8\,\mu s$, $P_{0}=2.5\,\textrm{mW}$]
   {    \psfrag{N}[c][c][0.9]{$N$}
    \psfrag{EE [Mbits/Joule]}[c][c][0.9]{$EE $ [Mbits/Joule]}
    \includegraphics[width=0.4\textwidth,height=0.2\textheight]{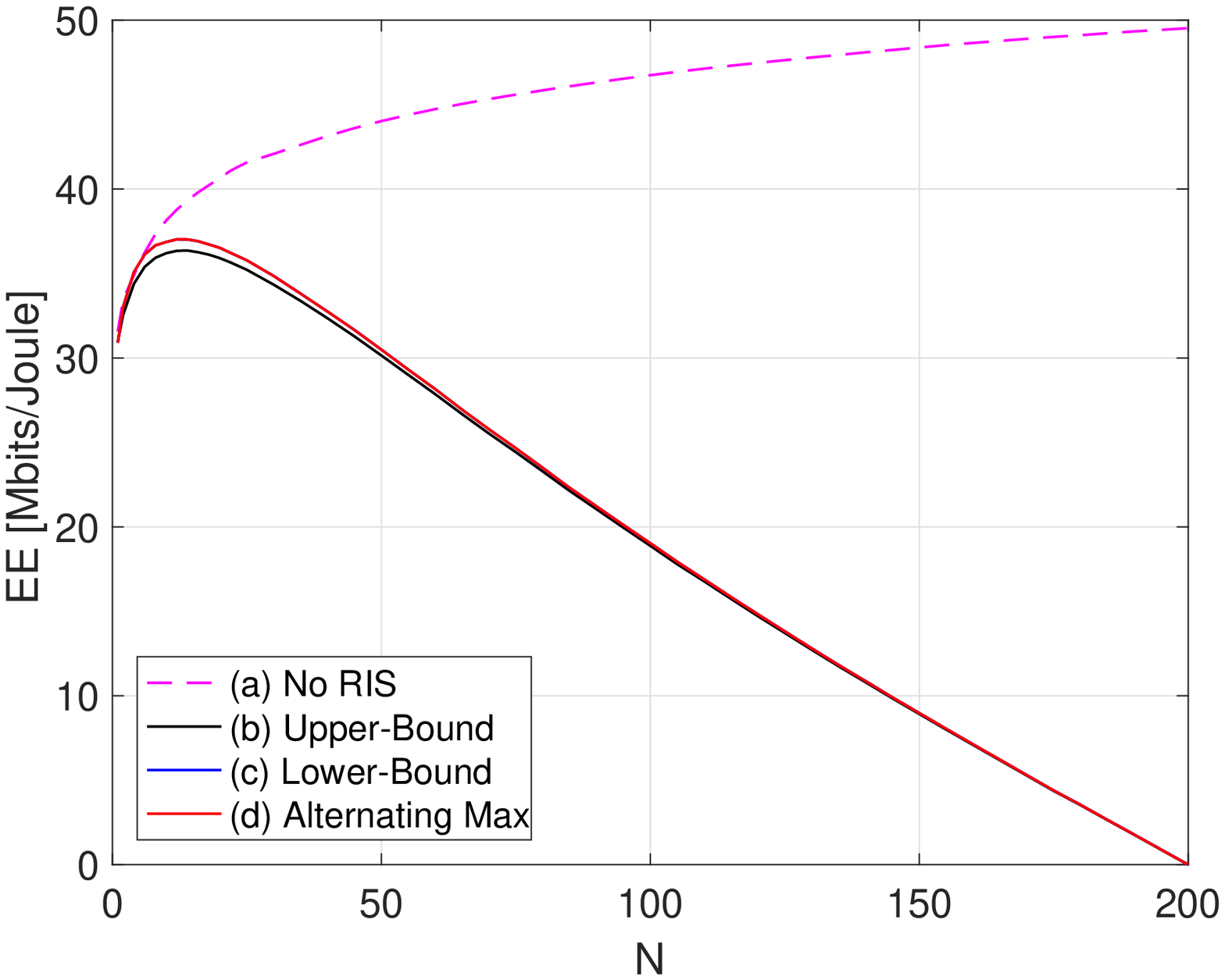}\label{fig:maxEE_NT8NR8_LargeT0}}
 \subfigure[EE for $T_{0}=0.15\,\mu s$, $P_{0}=0.5\,\textrm{mW}$]
   {    \psfrag{N}[c][c][0.9]{$N$}
    \psfrag{EE [Mbits/Joule]}[c][c][0.9]{$EE $ [Mbits/Joule]}
    \includegraphics[width=0.4\textwidth,height=0.2\textheight]{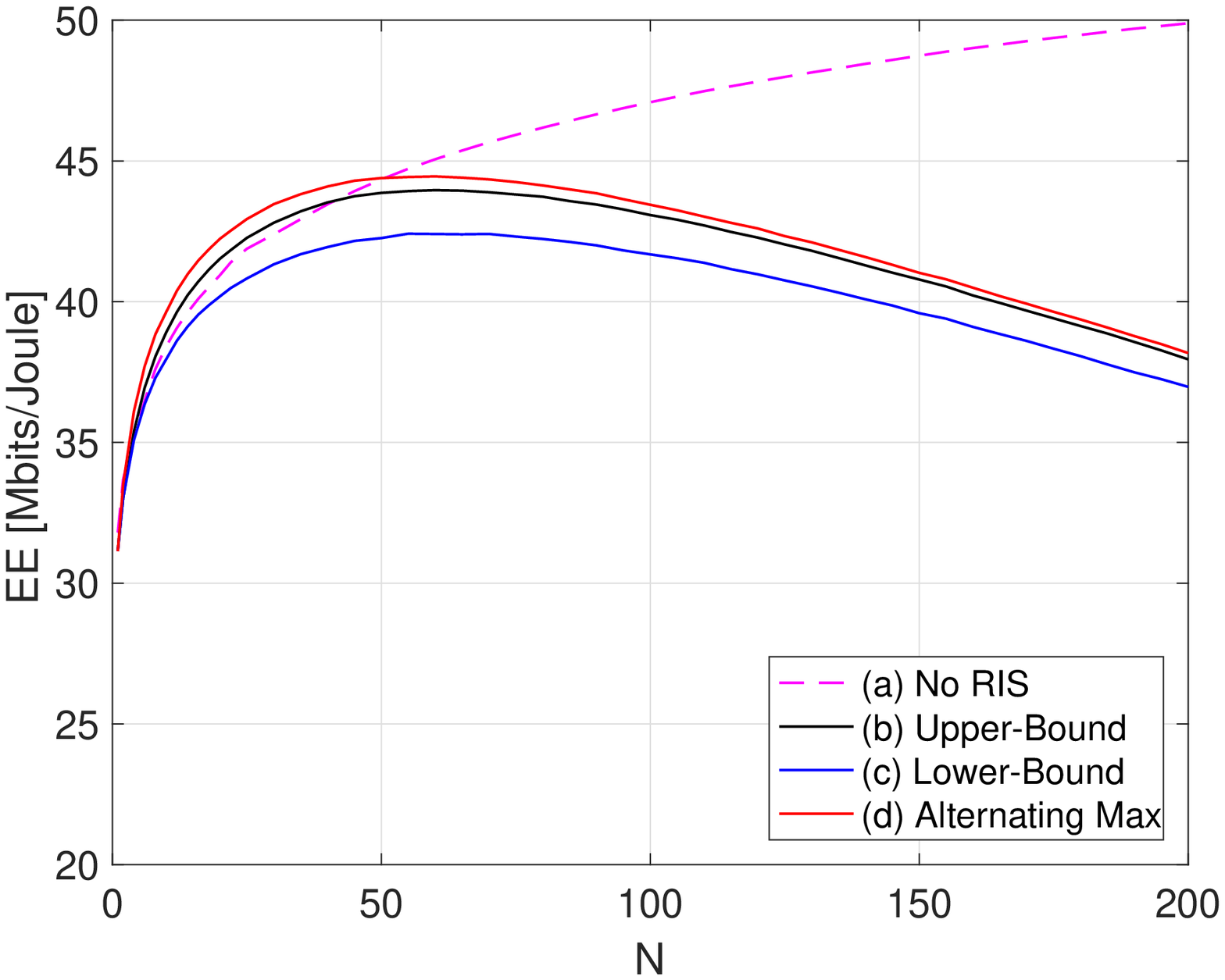} \label{fig:maxEE_NT8NR8_SmallT0}}
 \caption{Achieved EE in [Mbit/Joule] as a function of $N$ for $N_T=8$, $N_R=8$.}\label{fig:maxEE_NT8NR8}\vspace{-0.35cm}
 \end{figure}

\begin{figure}[!h]
 \centering
 \subfigure[EE vs. SE. $N=10$, $T_{0}=0.8\,\mu s$, $P_{0}=2.5\,\textrm{mW}$]
   {    \psfrag{SE [bits/s/Hz]}[c][c][0.9]{$SE$ [bits/s/Hz]}
    \psfrag{EE [Mbits/Joule]}[c][c][0.9]{$EE $ [Mbits/Joule]}
    \includegraphics[width=0.4\textwidth,height=0.2\textheight]{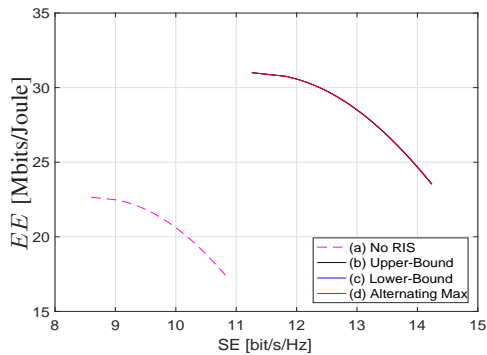}\label{fig:BiObj_NT1NR1_N10_LargeT0}}
 \subfigure[EE vs. SE $N=20$, $T_{0}=0.8\,\mu s$, $P_{0}=2.5\,\textrm{mW}$]
   {    \psfrag{SE [bits/s/Hz]}[c][c][0.9]{$SE$ [bits/s/Hz]}
    \psfrag{EE [Mbits/Joule]}[c][c][0.9]{$EE $ [Mbits/Joule]}
    \includegraphics[width=0.4\textwidth,height=0.2\textheight]{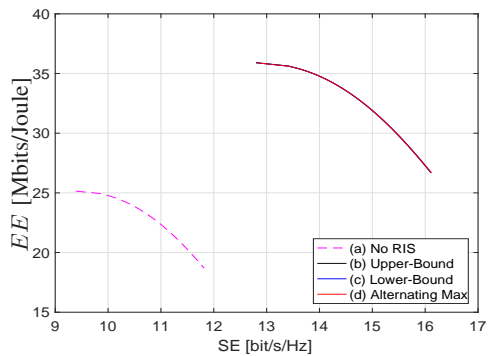} \label{fig:BiObj_NT1NR1_N20_LargeT0}}
 \subfigure[EE vs. SE. $N=50$, $T_{0}=0.8\,\mu s$, $P_{0}=2.5\,\textrm{mW}$]
   {    \psfrag{SE [bits/s/Hz]}[c][c][0.9]{$SE$ [bits/s/Hz]}
    \psfrag{EE [Mbits/Joule]}[c][c][0.9]{$EE $ [Mbits/Joule]}
    \includegraphics[width=0.4\textwidth,height=0.2\textheight]{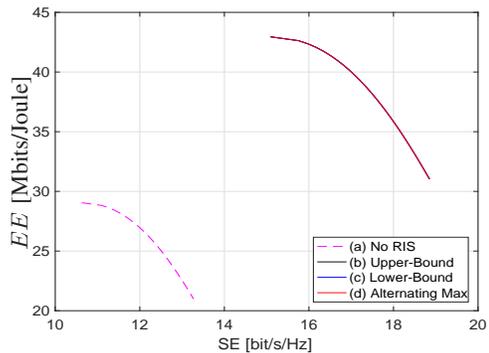} \label{fig:BiObj_NT1NR1_N50_LargeT0}}
\subfigure[EE vs. SE. $N=100$, $T_{0}=0.8\,\mu s$, $P_{0}=2.5\,\textrm{mW}$]
   {    \psfrag{SE [bits/s/Hz]}[c][c][0.9]{$SE$ [bits/s/Hz]}
    \psfrag{EE [Mbits/Joule]}[c][c][0.9]{$EE $ [Mbits/Joule]}
    \includegraphics[width=0.4\textwidth,height=0.2\textheight]{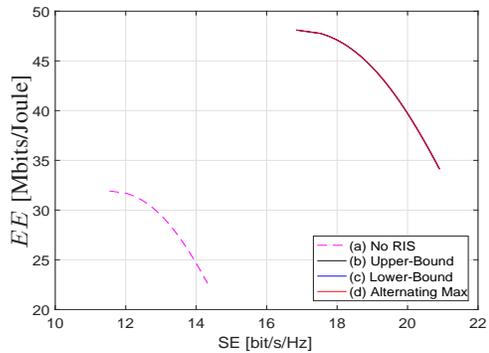} \label{fig:BiObj_NT1NR1_N100_LargeT0}}
 \caption{Achieved EE in [Mbit/Joule] as a function of achieved SE [bits/s/Hz] for $N_T=N_R=1$.}\label{fig:BiObj_NT1NR1_LargeT0} 
 \end{figure}

\begin{figure}[!h]
 \centering
 \subfigure[EE vs. SE. $N=10$, $T_{0}=0.8\,\mu s$, $P_{0}=2.5\,\textrm{mW}$]
  {    \psfrag{SE [Mbits/s]}[c][c][0.9]{$SE $ [bits/s/Hz]}
   \psfrag{EE [Mbits/Joule]}[c][c][0.9]{$EE$ [Mbits/Joule]}
   \includegraphics[width=0.4\textwidth,height=0.2\textheight]{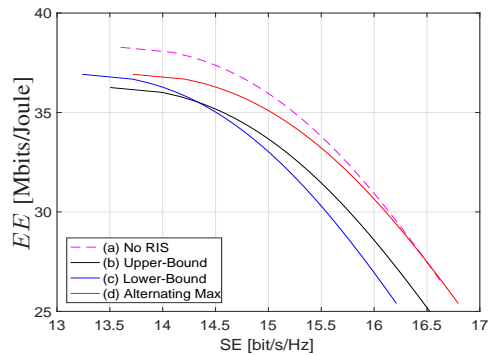}\label{fig:BiObj_NT8NR8_N10_LargeT0}}
 \subfigure[EE vs. SE. $N=20$, $T_{0}=0.8\,\mu s$, $P_{0}=2.5\,\textrm{mW}$]
   {    \psfrag{SE [bits/s/Hz]}[c][c][0.9]{$SE $ [bits/s/Hz]}
    \psfrag{EE [Mbits/Joule]}[c][c][0.9]{$EE $ [Mbits/Joule]}
    \includegraphics[width=0.4\textwidth,height=0.2\textheight]{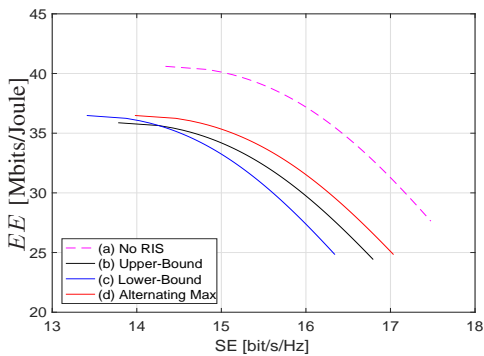} \label{fig:BiObj_NT8NR8_N20_LargeT0}}
 \subfigure[EE vs. SE. $N=50$, $T_{0}=0.8\,\mu s$, $P_{0}=2.5\,\textrm{mW}$]
   {    \psfrag{SE [bits/s/Hz]}[c][c][0.9]{$R $ [bits/s/Hz]}
    \psfrag{EE [Mbits/Joule]}[c][c][0.9]{$EE $ [Mbits/Joule]}
    \includegraphics[width=0.4\textwidth,height=0.2\textheight]{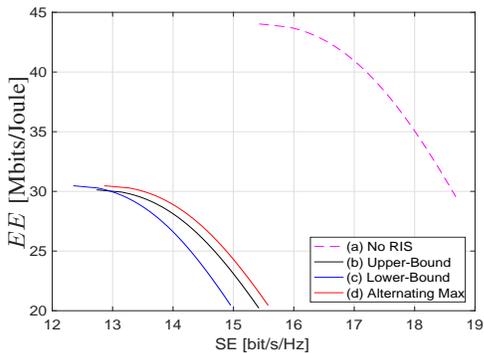} \label{fig:BiObj_NT8NR8_N50_LargeT0}}
\subfigure[EE vs. SE. $N=100$, $T_{0}=0.8\,\mu s$, $P_{0}=2.5\,\textrm{mW}$]
   {    \psfrag{SE [bits/s/Hz]}[c][c][0.9]{$R $ [bits/s/Hz]}
   \psfrag{EE [Mbits/Joule]}[c][c][0.9]{$EE $ [Mbits/Joule]}
    \includegraphics[width=0.4\textwidth,height=0.2\textheight]{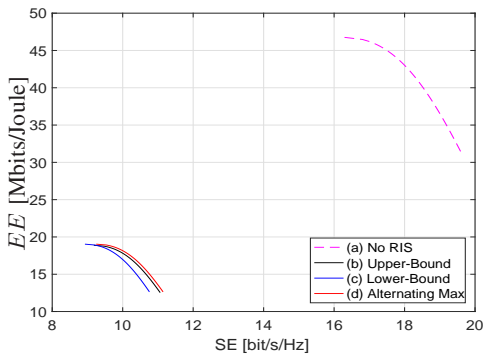} \label{fig:BiObj_NT8NR8_N100_LargeT0}}
 \caption{Achieved EE in [Mbits/Joule] as a function of achieved SE [bits/s/Hz] with $N_T=8$, $N_R=8$.}\label{fig:BiObj_NT8NR8_LargeT0}
\end{figure}

Similar considerations hold for the case in which the energy efficiency is optimized, as it emerges from Figures \ref{fig:maxEE_NT1NR1}, \ref{fig:maxEE_NT1NR8}, \ref{fig:maxEE_NT8NR8}, which consider the same four schemes considered in Figures \ref{fig:maxR_NT1NR1}, \ref{fig:maxR_NT1NR8}, \ref{fig:maxR_NT8NR8}, respectively, with the only differences that $p,p_{F},B,B_{F}$ have been allocated for energy efficiency maximization, according to the optimal method from Section \ref{Sec:EEOptimization}. Also, two values of $P_{0}$ are considered, namely $P_{0}=0.5\,\textrm{mW}$ and $P_{0}=2.5\,\textrm{mW}$. In this case, Scheme (a) without any \gls{ris} feedback transmission starts performing better than the optimized schemes that rely on feedback transmissions when $N_{R}=8$, $N_{T}=1$, $T_{0}=0.8\,\mu s$, and $N>150$, i.e., for a lower overhead than for rate optimization. This can be explained since in the case of energy efficiency optimization, feedback overheads do not affect only the rate function, but also the power consumption at the denominator of the energy efficiency in \eqref{Eq:aEEcont}. Finally, Figures \ref{fig:BiObj_NT1NR1_LargeT0} and \ref{fig:BiObj_NT8NR8_LargeT0} consider again Schemes (a)-(d), with $p,p_{F},B,B_{F}$ allocated for rate-energy bi-objective maximization according to the optimal method from Section \ref{Sec:BiObjectiveOpt}. The system rate-energy Pareto boundary is shown for the two cases: (1) $N_{T}=N_{R}=1$; (2) $N_{T}=N_{R}=8$, with $T_{0}=0.8\,\mu s$. Similar remarks as for previous scenarios hold.

\begin{figure}[!h]
 \centering
 \subfigure[$N_{T}=1$, $N_{R}=8$. SE for $T_{0}=0.8\,\mu s$]
   {    \psfrag{N}[c][c][0.9]{$N$}
    \psfrag{SE [bit/s/Hz]}[c][c][0.9]{$SE$ [bit/s/Hz]}
    \includegraphics[width=0.4\textwidth,height=0.2\textheight]{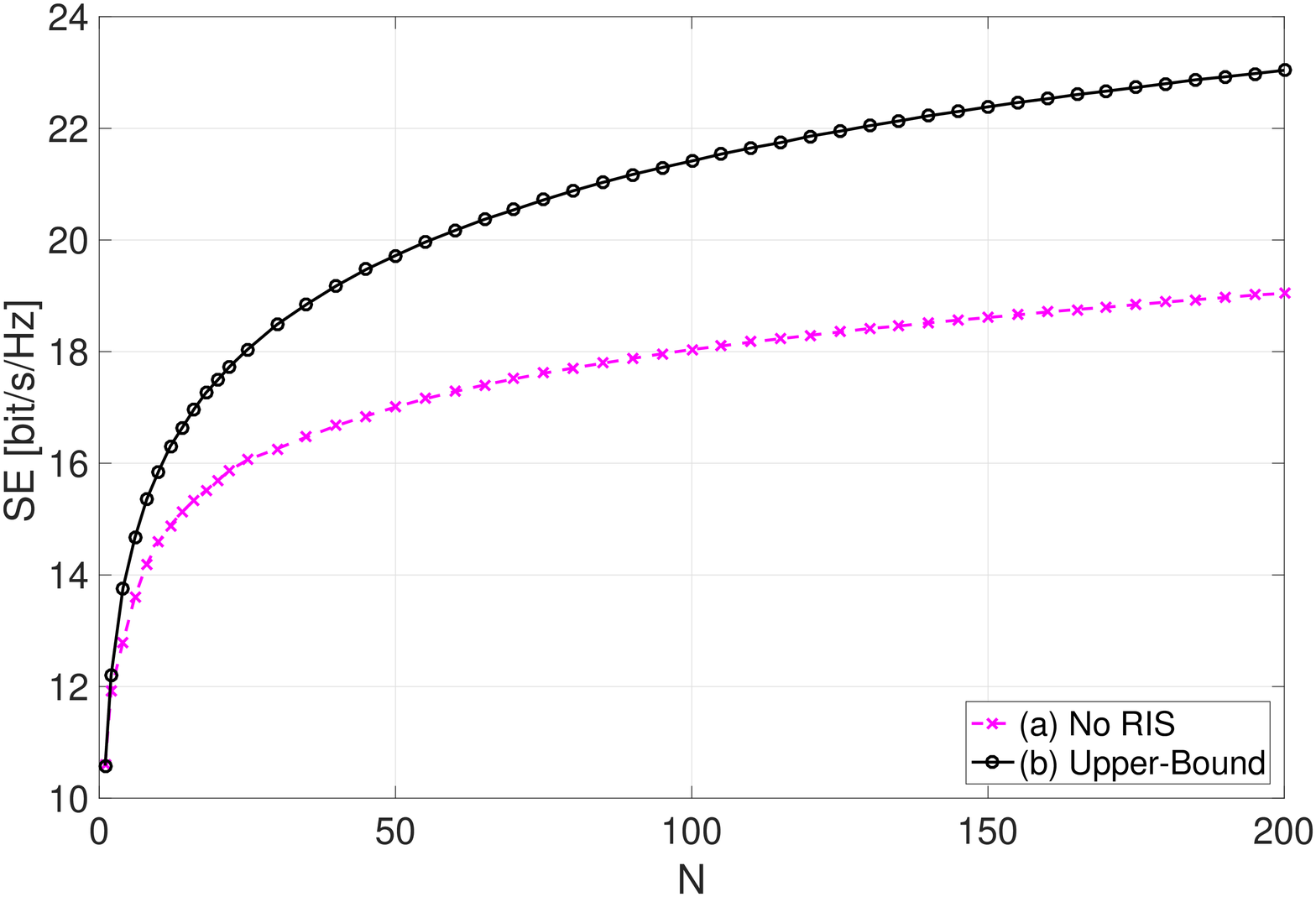}\label{fig:maxR_NT1NR8_LargeT0}}
\subfigure[$N_{T}=8$, $N_{R}=8$. SE for $T_{0}=0.8\,\mu s$]
   {    \psfrag{N}[c][c][0.9]{$N$}
    \psfrag{SE [bit/s/Hz]}[c][c][0.9]{$SE$ [bit/s/Hz]}
    \includegraphics[width=0.4\textwidth,height=0.2\textheight]{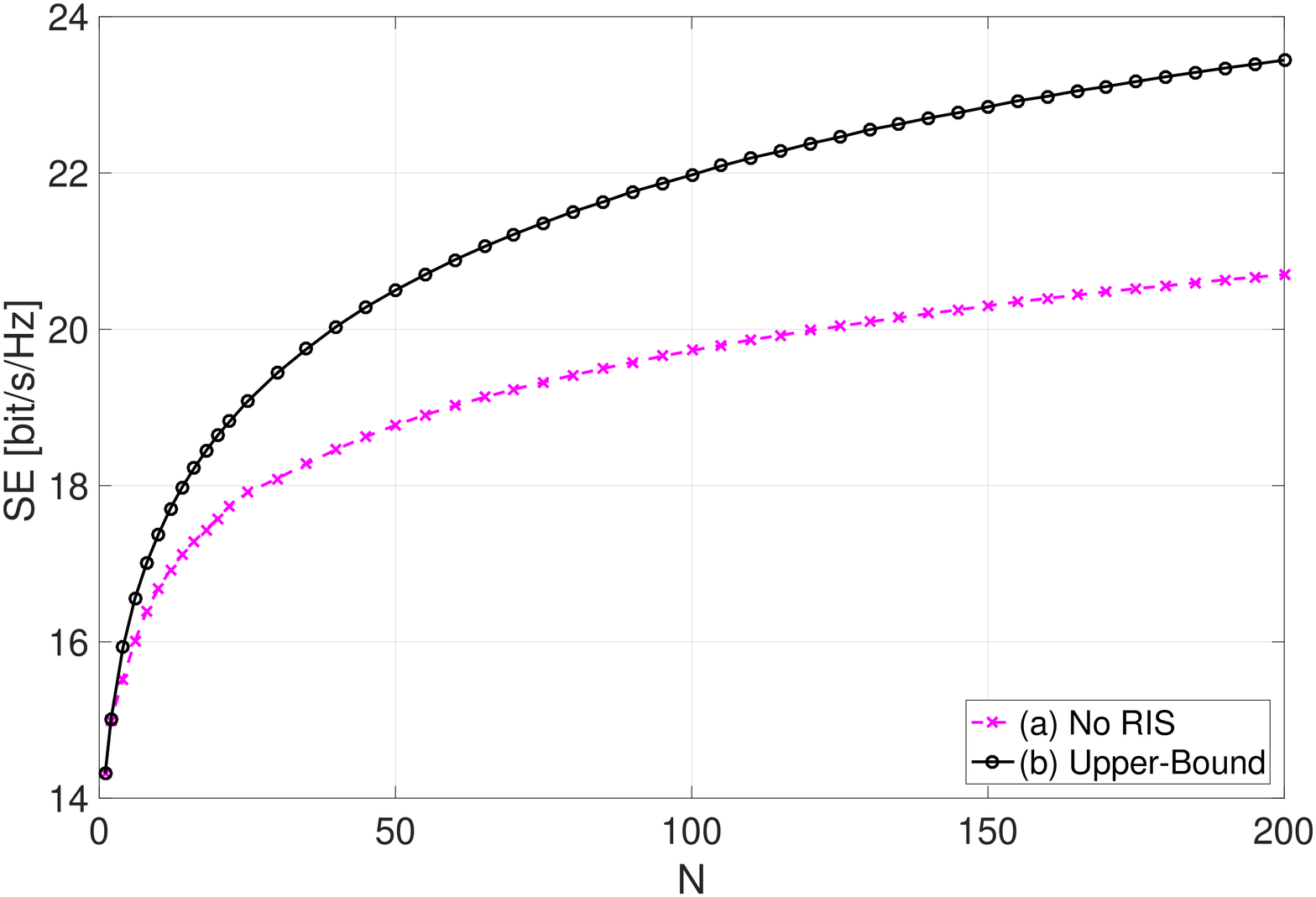}\label{fig:maxR_NT8NR8_LargeT0}}
\caption{SE in [bit/s/Hz] versus $N$ with TDD feedback}\label{fig:RateTDD}
\end{figure}

\begin{figure}
\subfigure[$N_{T}=1$, $N_{R}=8$. EE for $T_{0}=0.8\,\mu s$, $P_{0}=2.5\,\textrm{mW}$]
   {    \psfrag{N}[c][c][0.9]{$N$}
    \psfrag{EE [Mbit/J]}[c][c][0.9]{$EE$ [Mbit/J]}
    \includegraphics[width=0.4\textwidth,height=0.2\textheight]{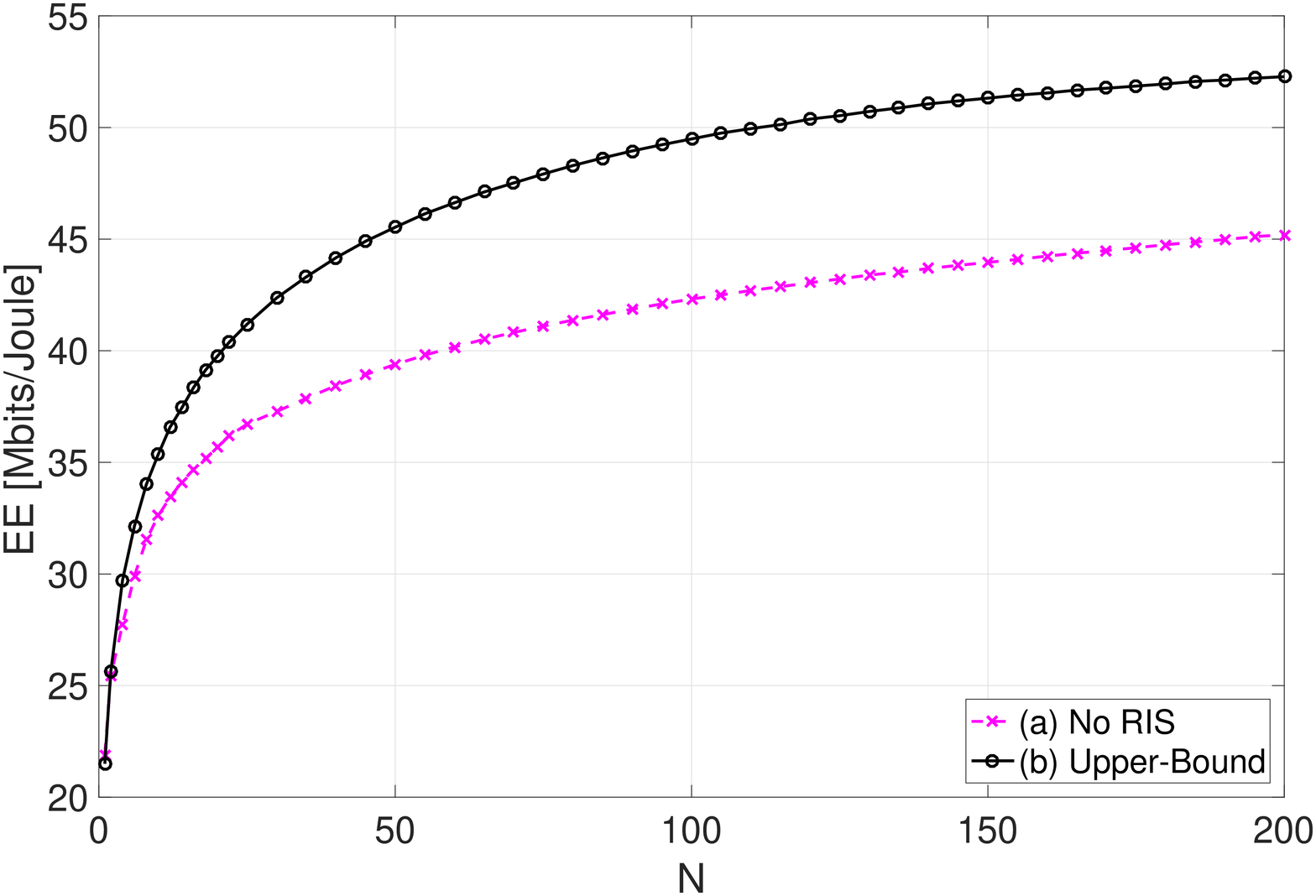}\label{fig:maxEE_NT1NR8_LargeT0}}
\subfigure[$N_{T}=8$, $N_{R}=8$. EE for $T_{0}=0.8\,\mu s$, $P_{0}=2.5\,\textrm{mW}$]
   {    \psfrag{N}[c][c][0.9]{$N$}
    \psfrag{EE [Mbit/J]}[c][c][0.9]{$EE$ [Mbit/J]}
    \includegraphics[width=0.4\textwidth,height=0.2\textheight]{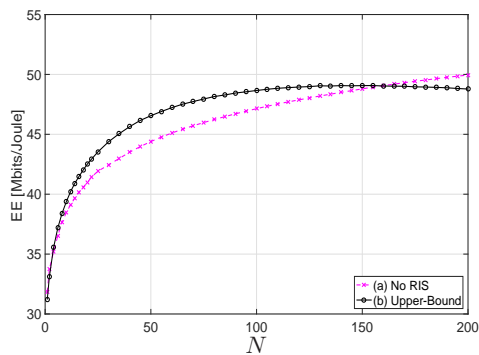}\label{fig:maxEE_NT8NR8_LargeT0}}
\caption{EE in [Mbit/Joule] versus $N$ with TDD feedback.}\label{fig:EE_TDD} 
\end{figure}

\begin{figure}
\centering
\subfigure[EE vs. SE. $N=20$, $T_{0}=0.8\,\mu s$, $P_{0}=2.5\,\textrm{mW}$]
 {    \psfrag{SE [bit/s/Hz]}[c][c][0.9]{$SE$ [bit/s/Hz]}
    \psfrag{EE [Mbit/Joule]}[c][c][0.9]{$EE$ [Mbit/Joule]}
   \hspace{-1.5cm} \includegraphics[width=0.4\textwidth,height=0.2\textheight]{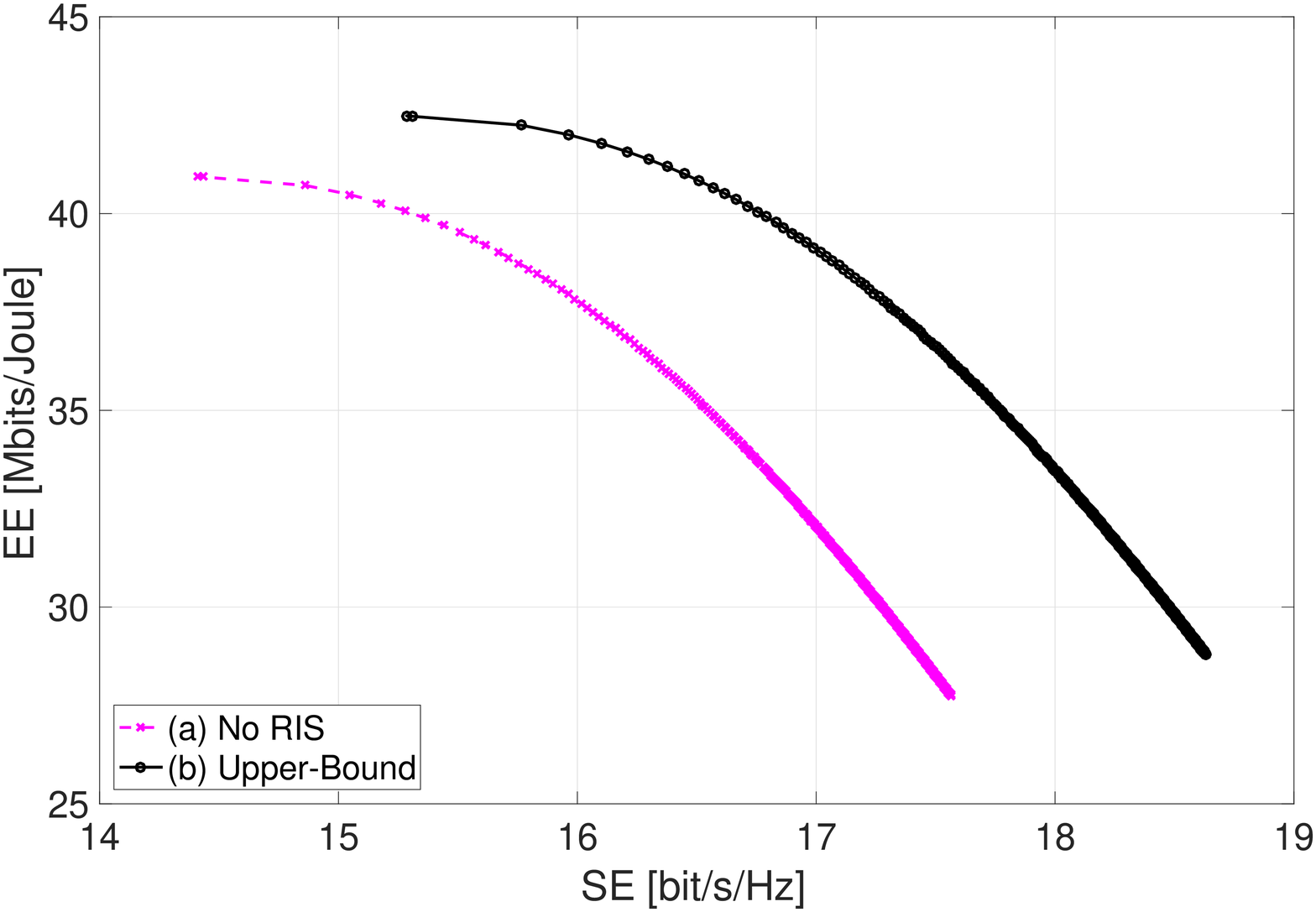} \label{fig:BiObj_NT8NR8_N20_LargeT0}}
\subfigure[EE vs. SE. $N=100$, $T_{0}=0.8\,\mu s$, $P_{0}=2.5\,\textrm{mW}$]
   {    \psfrag{SE [bit/s/Hz]}[c][c][0.9]{$SE$ [bit/s/Hz]}
    \psfrag{EE [Mbit/Joule]}[c][c][0.9]{$EE $ [Mbit/Joule]}
   \hspace{-1.5cm}  \includegraphics[width=0.4\textwidth,height=0.2\textheight]{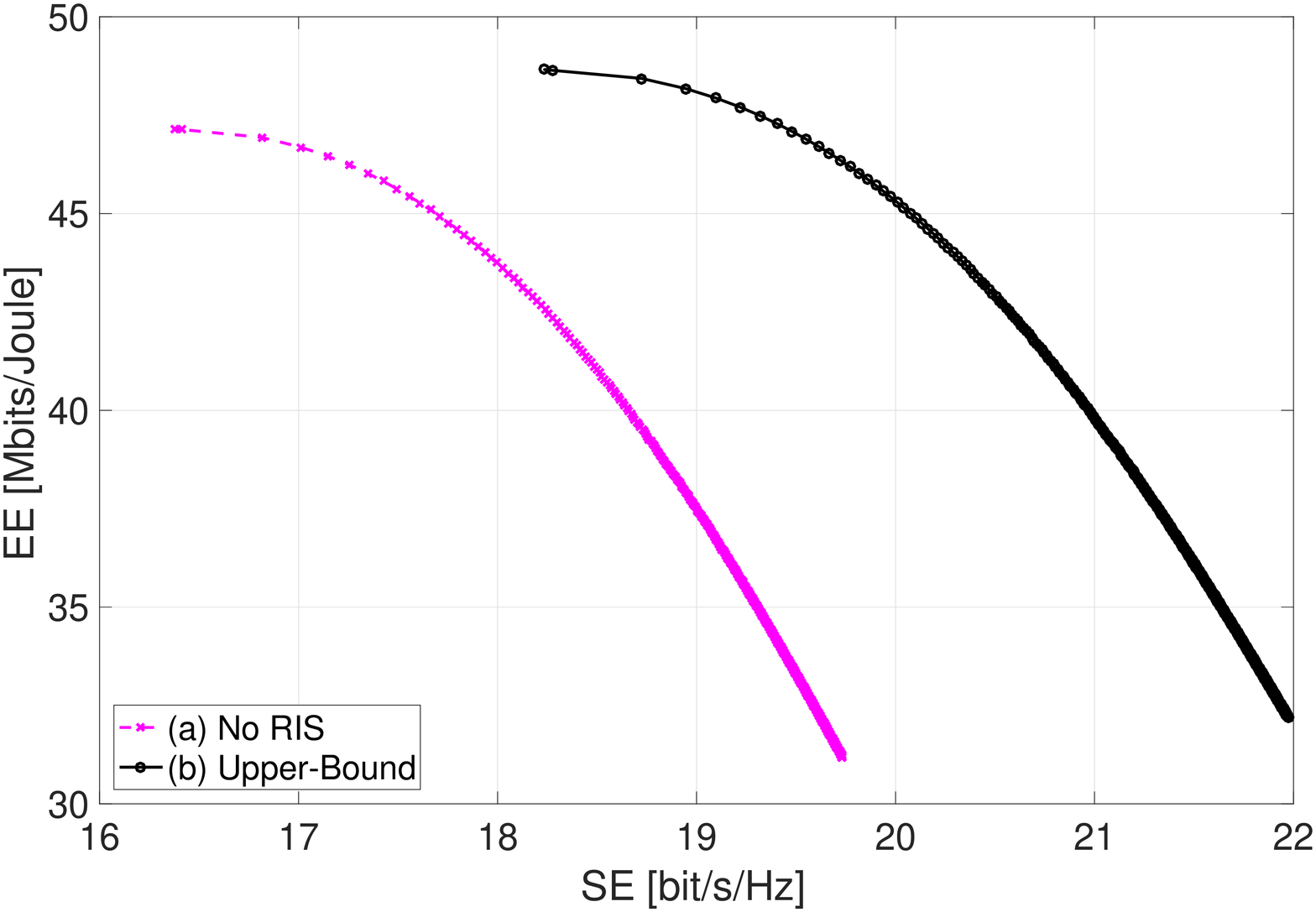} \label{fig:BiObj_NT8NR8_N100_LargeT0}}
\caption{EE [Mbit/Joule] versus SE [bit/s/Hz] with TDD feedback.}\label{fig:BiObjTDD} 
 \end{figure}

Next, Figures \ref{fig:RateTDD}, \ref{fig:EE_TDD}, \ref{fig:BiObjTDD} consider the overhead model in which the receiver transmit $N_{R}$ orthogonal pilots at the same time, as described in Section \ref{Sec:Overhead}, and show the achieved spectral efficiency, energy efficiency, and their optimal trade-off, for the case without \gls{ris} (Scheme (a)) and the use of Scheme (b) (similar results are obtained for Schemes (c) and (d), but results are omitted for brevity). Only the case $T_{0}=0.8\,\mu s$ and $P_{0}=2.5\,\textrm{mW}$ is considered, as this is the most difficult scenario for the proposed method since a longer time and more power are spent for each pilot tone. Despite the challenging scenario, Figure \ref{fig:RateTDD} shows that the slightly more sophisticated feedback scheme ensures that the use of an optimized \gls{ris} provides higher spectral efficiency in both cases $N_{R}=8\;,\;N_{T}=1$ and $N_{T}=N_{R}=8$. Similar results are shown in  Figure \ref{fig:EE_TDD} for the energy efficiency, with the difference that \gls{ris} optimization becomes not convenient when $N_{T}=N_{R}=8$ and $N\geq 150$, since transmitting the pilots simultaneously does not remove the factor $N_{R}$ in the term $P_{E}$. Finally, Figure \ref{fig:BiObjTDD} shows that \gls{ris} optimization improves the spectral energy trade-off when $N_{T}=N_{R}=8$ (and thus also when $N_{T}=1\,,\,N_{R}=8$), for $N=20$ and $N=100$.


\section{Conclusions}\label{Sec:Conclusions}
A framework for overhead-aware radio resource allocation in RIS-aided systems has been developed for spectral and energy efficiency optimization. Two new closed-form methods for the optimization of the \gls{ris} phase shifts, as well as of the transmit and receive vectors, have been developed. Moreover, the transmit powers and bandwidths for the communication and feedback phases have been globally optimized through concave/pseudo-concave maximizations. The derived results indicate that \gls{ris} constitutes a suitable technology when suitable feedback mechanisms are used or when few transmit and receive antennas are deployed, since a trade-off exists between optimizing the network radio resources and the overhead due to the deployment of the optimized solution. In particular, there exists a limit to the number of antennas and \gls{ris} reflectors, before feedback overhead makes radio resource optimization not convenient compared to the setup where \glspl{ris} are not deployed.
An important future line of investigation is the analysis of the impact of multi-user interference on overhead-aware resource allocation in \gls{ris}-based networks. Multi-user interference complicates the resource allocation problems, possibly requiring the use of numerical optimization techniques.

\bibliographystyle{IEEEtran}
\bibliography{FracProg}
\end{document}